\documentclass[12pt]{article}
\usepackage{amsmath,amssymb} 
\usepackage{amsfonts}
\usepackage{float}
\usepackage{graphicx}
\usepackage{subfig}
\usepackage[utf8]{inputenc} 
\usepackage{enumerate}
\usepackage{ytableau}
\usepackage{hyperref}
\usepackage{caption}
\usepackage{cite}
\usepackage{bm}
\usepackage{tikz}
\usepackage{blkarray}  
\usepackage{verbatim}
\begin{document}

\def\im{\text{i}}
\def\eqa{\begin{eqnarray}}
\def\eqae{\end{eqnarray}}
\def\be{\begin{equation}}
\def\ee{\end{equation}}
\def\bea{\begin{eqnarray}}
\def\eea{\end{eqnarray}}
\def\ba{\begin{array}}
\def\ea{\end{array}}
\def\bd{\begin{displaymath}}
\def\ed{\end{displaymath}}
\def\eg{{\it e.g.~}}
\def\ie{{\it i.e.~}}
\def\Tr{{\rm Tr}}
\def\tr{{\rm tr}}
\def\>{\rangle}
\def\<{\langle}
\def\a{\alpha}
\def\b{\beta}
\def\c{\chi}
\def\del{\delta}
\def\e{\epsilon}
\def\f{\phi}
\def\vf{\varphi}
\def\tvf{\tilde{\varphi}}
\def\g{\gamma}
\def\h{\eta}
\def\j{\psi}
\def\k{\kappa}
\def\l{\lambda}
\def\m{\mu}
\def\n{\nu}
\def\w{\omega}
\def\p{\pi}
\def\q{\theta}
\def\r{\rho}
\def\s{\sigma}
\def\t{\tau}
\def\u{\upsilon}
\def\x{\xi}
\def\z{\zeta}
\def\D{\Delta}
\def\F{\Phi}
\def\G{\Gamma}
\def\J{\Psi}
\def\L{\Lambda}
\def\W{\Omega}
\def\P{\Pi}
\def\Q{\Theta}
\def\S{\Sigma}
\def\U{\Upsilon}
\def\X{\Xi}
\def\nab{\nabla}
\def\pa{\partial}
\newcommand{\lra}{\leftrightarrow}

\newcommand{\bc}{{\mathbb{C}}}
\newcommand{\br}{{\mathbb{R}}}
\newcommand{\bz}{{\mathbb{Z}}}
\newcommand{\bp}{{\mathbb{P}}}

\def\({\left(}
\def\){\right)}
\def\nn{\nonumber \\}

\newcommand{\red}{\textcolor[RGB]{255,0,0}}
\newcommand{\blue}{\textcolor[RGB]{0,0,255}}
\newcommand{\green}{\textcolor[RGB]{0,255,0}}
\newcommand{\cyan}{\textcolor[RGB]{0,255,255}}
\newcommand{\magenta}{\textcolor[RGB]{255,0,255}}
\newcommand{\yellow}{\textcolor[RGB]{255,255,0}}
\newcommand{\sky}{\textcolor[RGB]{135, 206, 235}}
\newcommand{\orange}{\textcolor[RGB]{255, 127, 0}}
\def\d{\operatorname{d}}
\def\Renyi{R$\acute{\text{e}}$nyi }
\def\Poincare{Poincar$\acute{\text{e}}$ }
\def\Banados{Ba$\tilde{\text{n}}$ados }

\title{\textbf{Negative \Renyi entropy and Brane intersection}}
\vspace{14mm}
\author{Jia Tian$^1$\footnote{wukongjiaozi@ucas.ac.cn}\quad and \quad  Xiaoge Xu$^2$\footnote{xiaoge.xv@whu.edu.cn}
}
\date{}
\maketitle

\begin{center}
	{\it
		$^1$Kavli Institute for Theoretical Sciences (KITS),\\
		University of Chinese Academy of Science, 100190 Beijing, P.~R.~China\\
		$^2$School of Physics and Technology,\\
		Wuhan University, Wuhan, 430072, P.~R.~China
	}
\vspace{10mm}
\end{center}

\makeatletter
\def\blfootnote{\xdef\@thefnmark{}\@footnotetext}  % for blank footnote
\makeatother

\begin{abstract}
In this work, we revisit the calculation of \Renyi entropy in AdS$_3$/(B)CFT$_2$. We find that  gravity solutions brane intersection will lead to negative \Renyi entropy.
\end{abstract}

\baselineskip 18pt
\newpage

\tableofcontents

%%%%%%%%%%%%%%%%%%%%%%%%%%
\section{Introduction}
%%%%%%%%%%%%%%%%%%%%%%%%%%
\Renyi entropy \cite{renyi} which is a one-parameter generalization of von Neumann entanglement entropy is an important measure of entanglement in quantum field theories. It contains richer physical information. In principle knowing all the \Renyi entropies is equivalent to knowing all the eigenvalues of the reduced density matrix of a subsystem. In practice, \Renyi entropies are easier to numerically compute and experimentally measure.

\Renyi entropy has been extensively studied in many theories \cite{Franchini:2007eu,Hayden:2016cfa,Klebanov:2011uf,Holzhey:1994we,Calabrese:2009ez,Hartman:2013mia,Chen:2013kpa,Datta:2013hba,Perlmutter:2013paa,Perlmutter:2015iya,Headrick:2015gba,Perlmutter:2013gua,Lee:2014xwa,Hung:2014npa,Allais:2014ata,Lee:2014zaa,Lewkowycz:2014jia,Bueno:2015lza,Bianchi:2015liz,Dong:2016wcf,Headrick:2010zt,Hung:2011nu,Fursaev:2012mp,Faulkner:2013yia,Galante:2013wta,Belin:2013dva,Barrella:2013wja,Chen:2013dxa,Belin:2013uta,Nishioka:2013haa,Alday:2014fsa,Giveon:2015cgs,Chu:2016tps,Dey:2016pei,Belin:2017nze,Jiang:2017ecm,Dong:2018cuv,Donnelly:2018bef,Dong:2018lsk,Akers:2018fow,Rabenstein:2018bri,Jeong:2019ylz,Botta-Cantcheff:2020ywu,Dong:2020iod,Dong:2020uxp}. In this paper we focus on the \Renyi entropy in AdS$_3$/(B)CFT$_2$ \cite{Karch:2000gx,Takayanagi:2011zk}. Two-dimensional BCFT has an interesting quantity: the boundary entropy (or the logarithm of the g-function) \cite{Affleck:1991tk} which can also be computed holographically \cite{Azeyanagi:2007qj}. Even for two-dimensional CFTs without boundaries, the boundary entropy an important role.  \Renyi entropy can be defined as
\bea \label{renyientropy}
&&S_A^n=\frac{1}{1-n}\log \text{Tr}\rho_A^n\, , \label{renyientropy}\\
&&\text{Tr}\rho_A^n=\frac{Z_n(A)}{Z_1^n}\, , \label{rhon}
\eea 
where $A$ is a subsystem and $Z_n(A)$ is the path integral on a $n$--sheeted Riemann space $\mathcal{M}_n$ obtained by gluing $n$ copies of the original space $\mathcal{M}_1$ alone $ A$. The reason why boundary entropy arises is that \Renyi entropy has a UV singularity and to regularize it we can remove a small disk of radius $\epsilon$ around each branch point with conformal boundary conditions imposed on the circular boundary of the disk. Therefore the definition of \Renyi entropy depends on the regularization which is parameterized by conformal boundary condition or boundary entropy \cite{Cardy:2016fqc}. Not like the entanglement entropy is positive definite the boundary entropy can be both positive and negative. For example, the lower bound of boundary entropy of the unitary minimal models with central charge $c_m=1-\frac{6}{(m+1)m}$ is \cite{DiFrancesco:1997nk}
\bea
\frac{1}{4}\log\(\frac{8}{m(m+1)}\sin^2(\frac{\pi}{m})\sin^2(\frac{\pi}{m+1})\),
\eea 
which approaches $-\infty$ when $m\rightarrow \infty$.
Even though from the analysis of RG flow, one can argue that boundary entropy should have a lower bound \cite{Friedan:2005dj,Friedan:2012jk}. One may still wonder can boundary entropy is so negative that the \Renyi entropy becomes negative\footnote{Usually we treat the boundary entropy as a small correction so this never happens.   
But from the point of view of AdS/BCFT the boundary entropy can be arbitrarily large}. 
We would like to explore this possibility in AdS$_3$/BCFT$_2$. The bulk theory is AdS Einstein gravity theory with End-Of-the-World (EOW) branes. We will also generalize it to allow the configurations of brane intersections or corners. Such generalization has been explored recently in \cite{Miyaji:2022cma,Biswas:2022xfw,Kusuki:2022ozk}. However previous studies are restricted in the thermal AdS geometry and global AdS geometry (with a conical defect in the center) in which the static EOW brane configurations play no roles in the \Renyi entropy calculation\footnote{In \cite{Biswas:2022xfw}, the authors also studied the ``disk branes" which can reproduce the boundary entropy terms. However, in the Lorentz signature, these branes are not static.}. To be complementary, we consider the BTZ black geometry and excited geometry \footnote{which is obtained from the \Banados map as defined below.}. We find a connection between the negativity of the \Renyi entropy and brane intersection. Using these new bulk solutions we can reproduce the field theory results but these geometries also contain some exotic features such as orientation inversion. An inverted orientation results in a negative Euclidean time evolution which is argued to be ill-defined \cite{Kontsevich:2021dmb,Witten:2021nzp}. However similar geometries are also encountered in a recent work \cite{Bah:2022uyz} for estimating global charge violation.

We consider two following examples: a single interval on the infinite line and the interval at the end of a semi-infinite line in the ground state. These two systems are basically the same in the sense: after we remove small disks around branch points both of these two-dimensional Euclidean manifolds can be transformed into an annulus or a cylinder \cite{Cardy:2016fqc}. The only difference is that for the BCFT system, only one of the boundaries is a physical boundary while for the CFT system, both of the two boundaries are cut-off boundaries. However, when we compute the partition function the cut-off boundary plays the same role as the physical boundary does. Holographically they both correspond to an EOW brane in the gravity theory.

Here we distinguish the annulus geometry and cylinder geometry  because their bulk dual geometries are different. In particular, when we consider the BCFT system the regularization we take in the field theory naturally provides a regularization scheme in the holographic calculation and it turns out this regularization is exactly the one recently proposed in \cite{Kusuki:2022ozk} so as a by-product our results explain the origin of their cut-off proposal.

This paper is organized as follows. In section \ref{fieldcal} we revisit the calculation of \Renyi entropy in field theory with an emphasis on boundary entropy contribution which is usually ignored. In section \ref{holo}, after introducing the gravity theory set-ups, we reproduce the field theory results in two different geometries. We show the subtleties of the gravity solutions with brane intersection and show they lead to negative \Renyi entropy.

\section{Field theory calculation}
\label{fieldcal}
Let us start by briefly reviewing the field theory calculation of the \Renyi entropy. Let the subsystem $A$ of a CFT be the interval $x\in [-L,L]$. The simplest way to compute \eqref{rhon} is converting it to a two-point function of primary twisted operators $\sigma_h$ with conformal dimension $h_n=\bar{h}_n=(c/24)(n-(1/n))$ \cite{Calabrese:2004eu}. Thus from the general formula of the two-point function of primary fields in two-dimensional CFT, we get \cite{Calabrese:2004eu}
\bea \label{rho1}
\text{Tr}\rho_A^n=c_n\langle \sigma_h^\epsilon(-L)\sigma^\epsilon_{\bar{h}}(L)\rangle=c_n\(\frac{2L}{\epsilon}\)^{-\frac{c}{6}(n-\frac{1}{n})},
\eea 
and the resulting \Renyi entropy is 
\bea \label{renyi1}
S_A^n=\frac{c}{6}\frac{1+n}{n}\log\frac{2L}{\epsilon}+g_A+g_B.
\eea 
where $g_{A,B}$ are the boundary entropies that are related to the constant $c_n$ but can not be determined by this method directly. To derive the boundary term one we can compute the partition functions.

\subsection{From cylinder partition function}

\label{cylinder}
The Euclidean surface $\mathcal{M}_1$ can be mapped to a cylinder via the conformal transformation
\bea 
w=f(x)=\log\(\frac{x+L}{L-x}\),
\eea 
such that the two cut-off disks at $x=\pm L$ are mapped to two ends located at $w=\pm \log (2L/\epsilon)$ of the cylinder. Applying the open-closed duality the conformal boundary conditions are translated to two boundary states $|A(B)\rangle$ and partition functions on the cylinder can be written as
\bea 
Z_1=\tilde{q}^{-c/24}\sum_k \langle A|k\rangle \langle k|B\rangle \tilde{q}^{\Delta_k},\quad Z_n=\tilde{q}^{-c/24n}\sum_k \langle A|k\rangle \langle k|B\rangle \tilde{q}^{\Delta_k/n},
\eea 
where the modular parameter is defined as $
\tilde{q}=e^{-4\pi W/\beta}
$ 
and  $\beta$ is the circumference of the cylinder.
In the limit of $W/\beta>>1$, only the ground state contributes
\bea 
\tr{\rho^n}\equiv \frac{Z_n}{Z_1^n}=\(\langle A|0\rangle \langle 0|B\rangle \)^{1-n}e^{(\frac{1}{n}-n)\frac{\pi W}{\beta}\frac{c}{6}},
\eea 
so the \Renyi entropy is
\bea \label{renyic}
S^{(n)}_A=\frac{n+1}{n}\frac{\pi c}{6\beta}W+g_A+g_B,
\eea 
where $g_{A,B}=\log\langle A,B|0\rangle$ are the boundary entropies. 

In our case, $\beta=2\pi$ and $W=2\log(2L/\epsilon)$ so the \Renyi entropy matches \eqref{renyi1}. The first \Renyi entropy $S_A^{(1)}$ \footnote{Strictly speaking, \Renyi entropy is only defined for $n\neq 1$ and $S_A^1$ should be defined by a proper analytic continuation.} is the entanglement entropy where the boundary entropy term can be absorbed into the UV cut-off $\epsilon$ then the result matches the standard result. However, it is pointed out in \cite{Cardy:2016fqc}, the boundary entropies are in principle measurable by considering different values of $n$. Therefore one should expect the dual holographic calculation should be able to reproduce them. The advantage of using cylinder geometry is the replicated  geometry $\mathcal{M}_n$ is also a cylinder so that the bulk dual is easy to find.
\subsection{From Weyl anomaly and Weyl transformation in annulus spacetime}
\label{annulus}

The regulated surface can also be mapped to an annulus via the conformal transformation
\bea 
f(x)=\frac{x+L}{L-x}.
\eea 
Let the metric of the annulus be 
\bea 
ds^2_{\mathcal{M}_1}=dr^2+r^2 d\theta^2.
\eea 
The  annulus partition function has a universal contribution from the Weyl anomaly due to the existence of boundaries \cite{Herzog:2015ioa}:
\bea \label{mm1}
I(\mathcal{M}_1)\sim -\frac{c}{6}\log\frac{r_{\text{max}}}{r_{\text{min}}}
\eea 
where $r_{\text{max}}=2L/\epsilon$ ($r_{\text{min}}=\epsilon/(2L)$) is the radius of the outer (inner) boundary. Thus the corresponding partition function is 
\bea \label{fz1}
Z_1\sim e^{-I(\mathcal{M}_1)}=e^{\frac{c}{3}\log\frac{2L}{\epsilon}}Q_\epsilon,
\eea 
where $Q$ is a constant that is regularization dependent \cite{Lunin:2000yv}.
The $n$-sheeted cover of the annulus has the metric 
\bea
ds^2_{\mathcal{M}_n}=dr^2+n^2r^2 d\theta^2=d\zeta d\bar{\zeta},
\eea 
where $\zeta=r e^{\im n \phi}$. Introducing $\xi=\zeta^{1/n}$ we can find that the it relates \eqref{mm1} via a Weyl transformation 
\bea 
ds^2_{{\mathcal{M}}_n}=n^2 \r^2(d\r^2+\r^2 d\phi^2)=e^{-2\tau}ds^2_{\hat{\mathcal{M}}_n},\quad e^{-\tau}=n \r^{n-1},\quad \xi=\r e^{\im \phi}.
\eea 
It is well known that the 2d Weyl anomaly is described by the Liouville action \cite{Liouville}
\bea 
\frac{Z[e^{-2\tau}g]}{Z[g]}=\exp \left[-\frac{c}{24\pi}\int_{\mathcal{M}} \sqrt{g}[R\tau-(\partial \tau)^2 ]d^2x-\frac{c}{12\pi}\int_{\partial \mathcal{M}} K \tau \sqrt{\gamma}dy\right],
\eea 
which implies 
\bea 
Z[\mathcal{M}_n]&=&Z[\hat{\mathcal{M}}_n]\exp\(\frac{c}{12}(n^2-1)\log\frac{\rho_{\text{max}}}{\rho_{\text{min}}}\) \\
&=&\exp\(\frac{c}{6}\log \frac{\rho_{\text{max}}}{\rho_{\text{min}}}+\frac{c}{12}(n^2-1)\log\frac{\rho_{\text{max}}}{\rho_{\text{min}}}\)Q_\epsilon\\
&=&\exp\(\frac{c}{12}(n^2+1)\log\frac{\rho_{\text{max}}}{\rho_{\text{min}}}\)=\exp\(\frac{c}{6}(n+\frac{1}{n})\log\frac{2L}{\epsilon}\)Q_\epsilon.\label{weyln}
\eea 
Therefore using \eqref{renyientropy} and $\rho_{\text{max(min)}}=r^{1/n}_{\text{max(min)}}$ we get the same result as \eqref{renyic}
\bea 
\text{Tr}\rho_A^n=\exp\(\frac{c}{6}(\frac{1}{n}-n)\log\frac{2L}{\epsilon}\)=\(\frac{2L}{\epsilon}\)^{-\frac{c}{6}(n-\frac{1}{n})}Q_\epsilon^{1-n},
\eea 
if we identify $Q_\epsilon=g_A g_B$. The identification is due to the fact the constant $Q$ is given by the path integral in the holes with state $|A,B\rangle$ thus it is equal to the inner product $\langle A,B|0\rangle$.

\section{Holographic calculation}
\label{holo}
\subsection{set-ups}
\label{set-ups}
For the holographic calculation, we consider the Euclidean gravity action on a manifold $M$ potentially with a corner. We assume that the corner is bounded by two codimension-1 surfaces $\Sigma_1$ and $\Sigma_2$ and the corner is a codimension-2 surface $\Gamma$. The full action is given by
\bea \label{gravity}
I_M=-\frac{1}{16\pi G_N}\int_M \sqrt{g}(R-2\Lambda)-\frac{1}{8\pi G_N}\int_{\Sigma_1,\Sigma_2}\sqrt{h}(K-T)+\frac{1}{8\pi G_N}\int_\Gamma(\theta-\theta_0)\sqrt{\gamma},
\eea 
where the last term is the analog of the Hayward term \cite{Hayward:1993my} and $\theta_0$ is a fixed value that characterizes the corner.
The two codimension-1 surfaces can be AdS asymptotic boundaries, hard cut-off surfaces, or EOW boundaries with tension $T$. On $\Sigma_{1,2}$ we need to impose proper boundary conditions such that the variation of the action is well-defined. The variation of the action $I_M$ is given by
\bea 
&&\delta I_M= -\frac{1}{16\pi G_N}\int_M \(R_{\mu\nu}-\frac{1}{2}Rg_{\mu\nu}+\Lambda g_{\mu\nu}\)\delta g^{\mu\nu}\nonumber \\ &&\qquad-\frac{1}{8\pi G_N}\int_{\Sigma_{1,2}}\sqrt{h}(K_{ab}-h_{ab}(K-T))\delta h^{ab}
+\frac{1}{8\pi G_N}\int_\Gamma (\theta-\theta_0)\delta \sqrt{\gamma}. \label{va}
\eea 
If we choose the Dirichlet boundary condition on $\Sigma_{1,2}$   i.e. $\delta\Sigma_{1,2}=0 $, then $\delta \sqrt{\gamma}$ is also fixed thus the variation is well defined. However,  if we choose the Neumann boundary condition:
\bea 
K_{ab}-h_{ab}(K-T)=0,\quad h^{ab}~ \text{is free},\label{beom}
\eea 
there will be two possible choices $I:\theta-\theta_0=0$ or $ II: \delta \sqrt{\gamma}=0$ such that the last term \eqref{va} vanishes. To reproduce the field theory result we take the first choice and treat $\theta_0$ as the parameter of the corner determined by the equation $\theta-\theta_0=0$. For this choice, the corner term will not contribute to the on-shell action directly and the gravity theory can be viewed as a generalization of standard AdS/BCFT.

\subsection{Cylinder boundary}
\begin{figure}[h!]
	\centering
	\subfloat[]{
		\centering
		\includegraphics[scale=0.25]{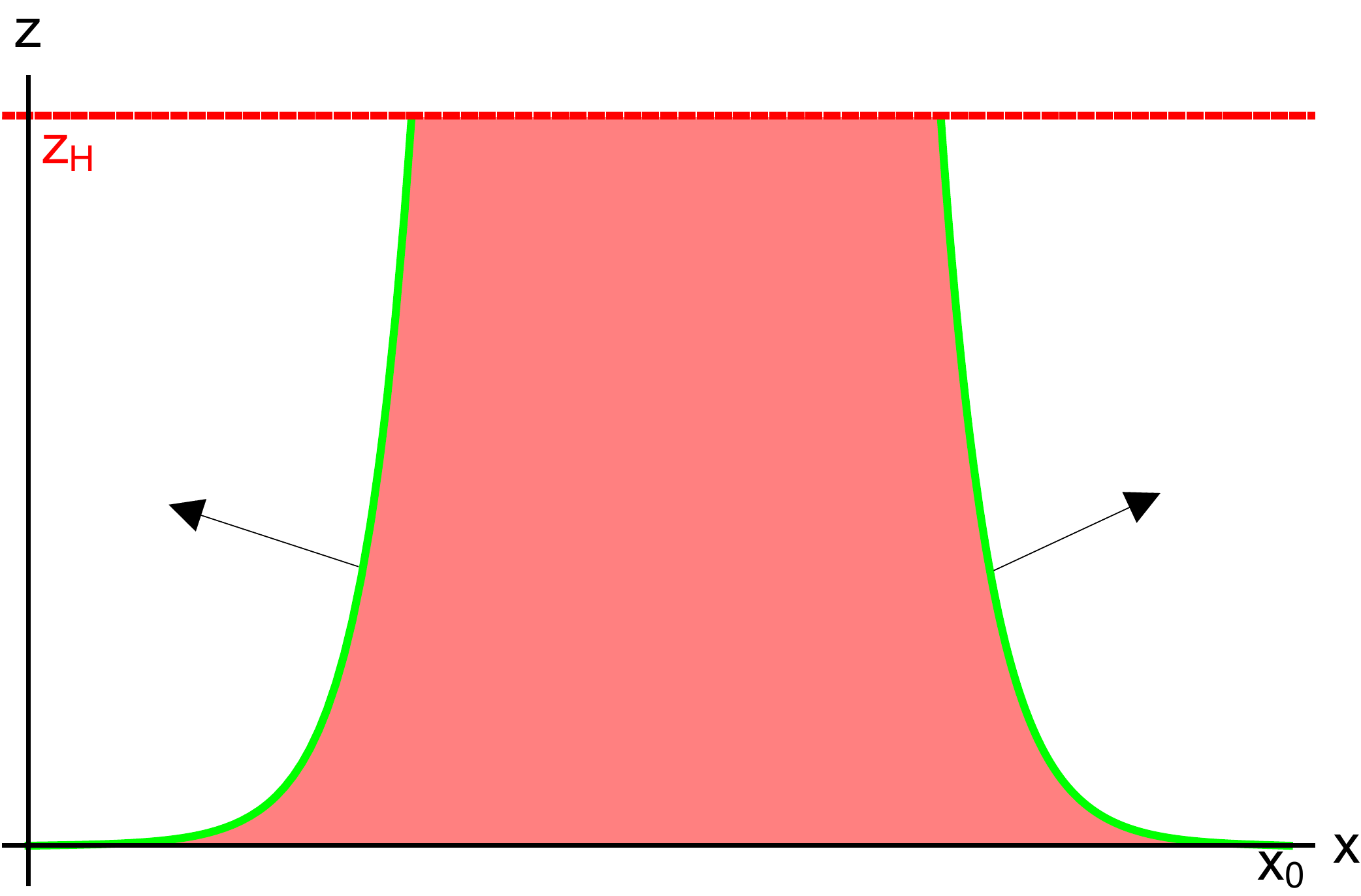}}
	\hfill
	\subfloat[]{
		\centering
		\includegraphics[scale=0.25]{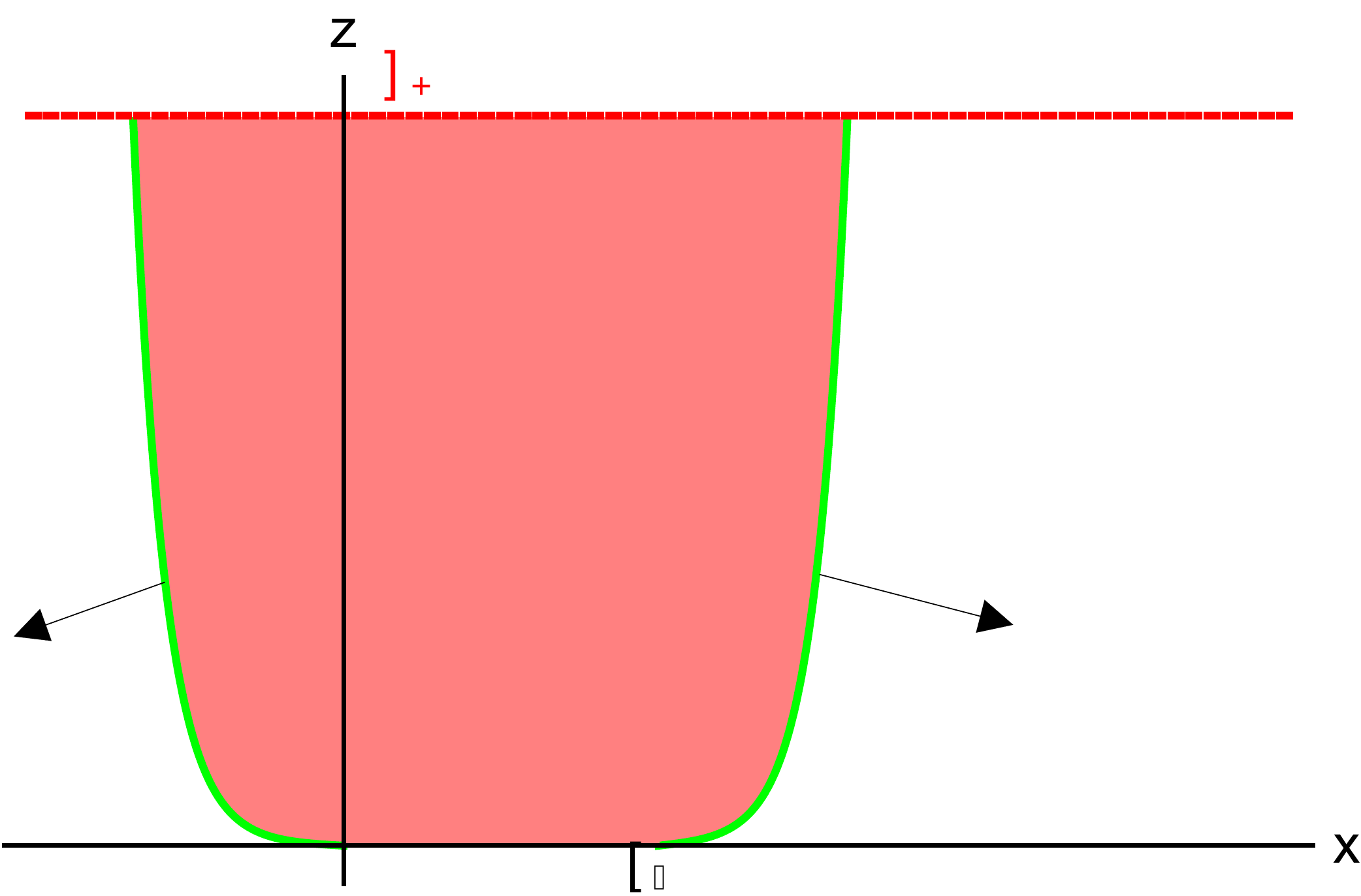}}
	
	\subfloat[]{
		\centering
		\includegraphics[scale=0.25]{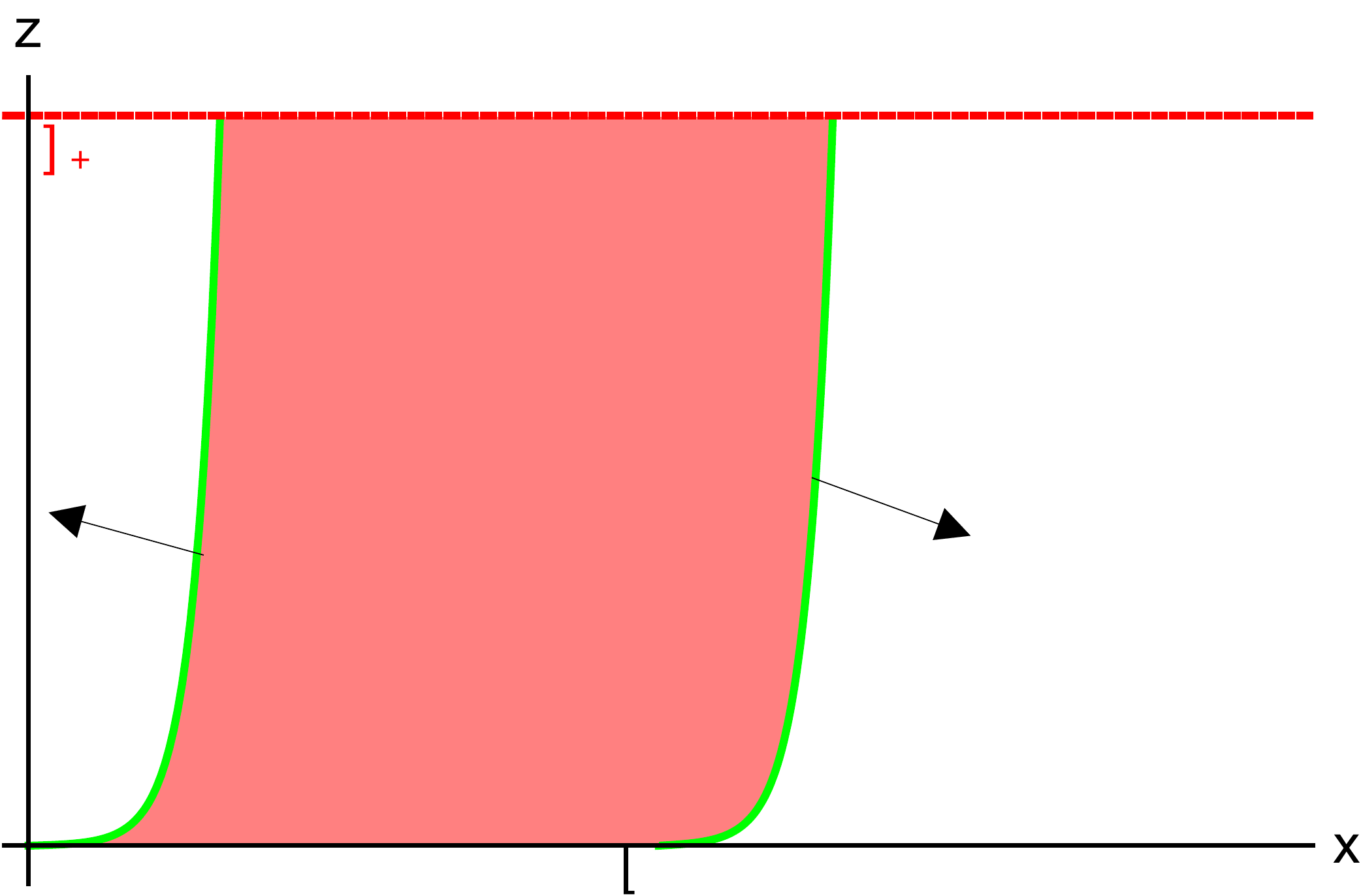}}
	\hfill
	\subfloat[]{
		\centering
		\includegraphics[scale=0.25]{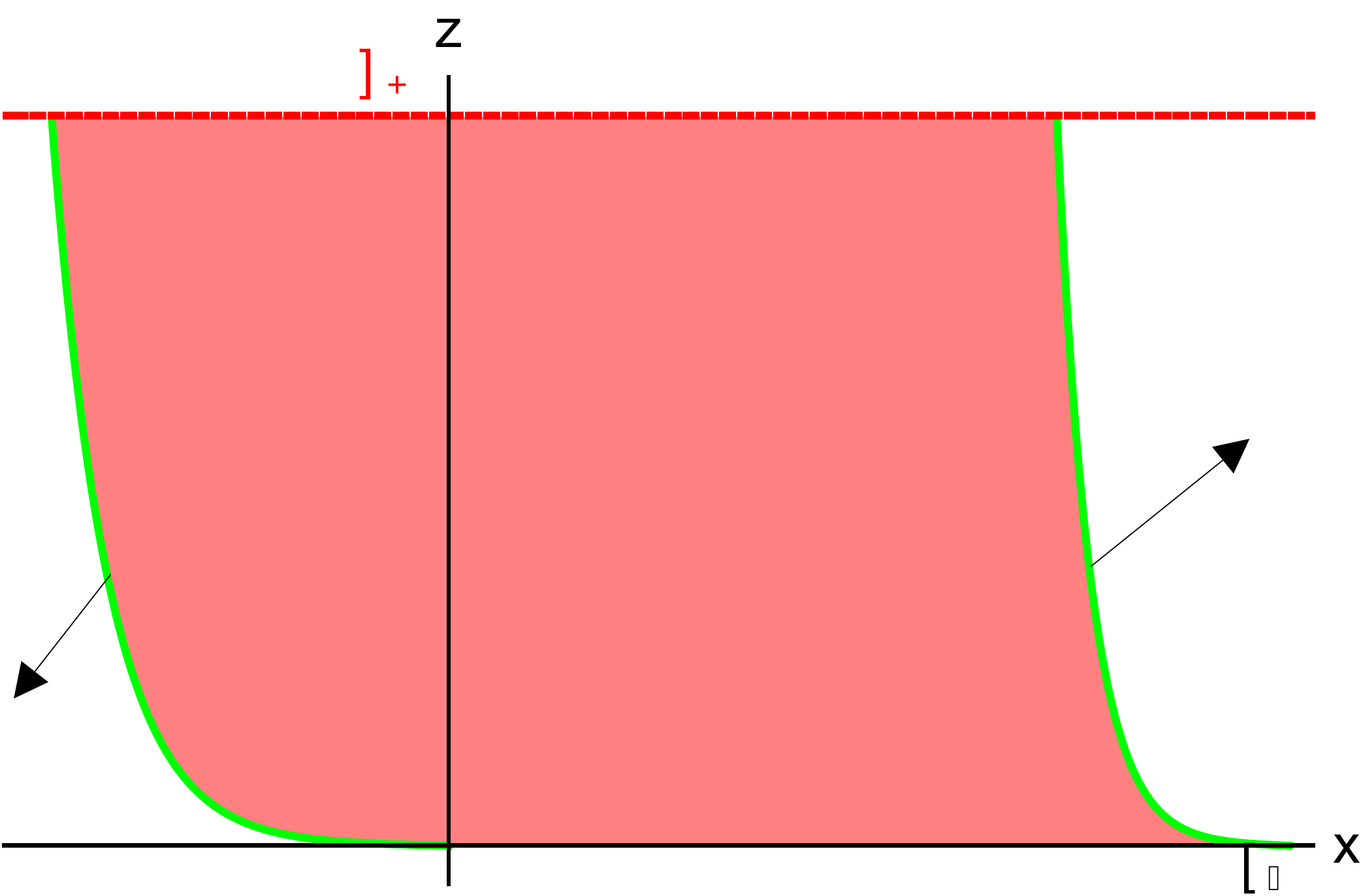}}
	\caption{Since each EOW brane can have either positive or negative brane tension then there are in total four possible configurations. When the normal vector points in the positive (negative) $z$ direction the brane tension is negative (positive).}\label{2eows}
\end{figure}

When the boundary BCFT is defined on a cylinder geometry, the bulk solution can be thermal AdS or BTZ black holes and there is also a Hawking-Page-like phase transition.   
Note that the length of the cylinder is very long $2\log(2L/\epsilon)$ then it should be in the high-temperature phase in the open channel. Thus to reproduce the \Renyi entropy we should consider the BTZ black hole solutions. The thermal AdS and global AdS with intersecting EOW branes have been studied in \cite{Miyaji:2022cma}. Indeed for these two cases, the on-shell action will not depend on the details of the  static EOW branes so boundary entropy terms in the \Renyi entropy can not be reproduced. The metric of the (non-rotating) BTZ black hole is 
\bea \label{BTZ}
ds^2=\frac{1}{z^2}\({f(z)} d \tau^2+\frac{\d z^2}{ f(z)}+ dx^2\),
\eea 
where  $f(z)=1-(z/z_H)^2,\tau\sim \tau+2\pi z_H$. In the BTZ geometry, the profile of the static EOW brane is \cite{Fujita:2011fp}
\bea 
Q_\pm(x_0,\lambda):\quad z= \pm \frac{z_H}{\lambda}\sinh \frac{x-x_0}{z_H},\quad \lambda=|\frac{T}{\sqrt{1-T^2}}|,\label{lambda}
\eea 
which ends on the AdS boundary at $x=x_0$. If we assume that the BCFT is defined on a cylinder then we need to insert two EOW branes in the bulk as shown in Fig.\ref{2eows} 
The on-shell action of configurations $(b)$ has been computed in \cite{Fujita:2011fp}. For other cases, the calculation is almost the same here as an illustration we present the calculation for configurations $(c)$ where the two EOW branes are parameterized as
\bea \label{ceow1}
Q_1: z=\frac{z_H}{\lambda_1}\sinh \frac{x}{z_H},\quad Q_2: z=\frac{z_H}{\lambda_2}\sinh \frac{x-x_0}{z_H},\quad x_0>0.
\eea 
The bulk contribution to the action is 
\bea 
I_{\text{bulk}}&=&-\frac{1}{16\pi G}\int (R-2\Lambda)\sqrt{g} \\&=&\frac{\beta}{4\pi G}\(\int_{\epsilon}^{z_1}\frac{1}{z^3}dz\int_0^{x_1}dx+\int_{z_2}^{z_H}\frac{1}{z^3}dz\int^{x_2}_{x_0}dx+\int_\epsilon^{z_H}\frac{1}{z^3}dz\int_{x_1}^{x_0}dx\)\\
&=&\frac{\beta}{8\pi G}\(-\frac{x_2-x_1}{z_H^2}-\int_0^{x_1}\frac{1}{z_1^2}dx+\int_{x_0}^{x_2}\frac{1}{z_2^2}dx+\frac{x_0}{\epsilon^2}\),
\eea 
the boundary contribution is
\bea 
I_{\text{bdy}}=-\frac{1}{8\pi G}\int \sqrt{h}(K-T)=-\frac{\beta}{8\pi G}\(-\int_0^{x_1}\frac{1}{z_1^2}dx+\int_{x_0}^{x_2}\frac{1}{z_2^2}dx\).
\eea 
and the counter term is 
\bea 
I_{ct}&=&-\frac{1}{8\pi G}\int \sqrt{\gamma}\label{ct}\\
&=&-\frac{\beta x_0 }{8\pi G} (\frac{1}{\epsilon^2}-\frac{1}{2z_H^2}),
\eea 
where $\gamma_{mn}$ is the induced metric on the cut-off surface $z=\epsilon$. Adding these terms together we obtain the final regulated on-shell action
\bea 
I_{\text{on-shell}}(z_H)&=&\frac{1}{4 G}\(\frac{x_0}{2z_H}-\frac{x_2-x_1}{z_H}\)=\frac{1}{4 G}\(-\frac{x_0}{2z_H}+\frac{x_1}{z_H}-\frac{x_2-x_0}{z_H}\)\\
&=&-\frac{c}{12}\frac{2\pi}{\beta}W+\frac{c}{6}\sinh ^{-1}(\lambda_1 ) -\frac{c}{6}\sinh ^{-1}(\lambda_2 ). \label{expect}
\eea 
Noticing that the bulk dual of $\mathcal{M}_n$ is also  described by \eqref{BTZ} with the replacement $z_H\rightarrow n z_H$ then we can reproduce the \Renyi entropy 
\bea 
S_A^{(n)}&=&\frac{1}{1-n}\log \tr \rho_A^n=\frac{1}{1-n}\log \frac{e^{-I_{\text{on-shell}}(nz_H)}}{e^{-n I_{\text{on-shell}}(z_H)}}=\frac{c}{12}\frac{1+n}{n}W+g_A+g_B,\label{entropyc}\\
&=&\frac{c}{6}\frac{1+n}{n}\log\frac{2L}{\epsilon}+g_A+g_B,\label{renyi2}
\eea 
where we have used the relation $W=2\log\frac{2L}{\epsilon}$ and identification $g_A=-\frac{c}{6}\sinh ^{-1}(\lambda_1 ),g_B=\frac{c}{6}\sinh ^{-1}(\lambda_2 )$. One of the observations we want to make is that the EOW brane with negative tension has a negative contribution to the \Renyi entropy! The other observation is that if one of the branes has negative tension (configurations $(a),(b),(c)$) then it is possible that the two branes will intersect outside of the horizon. So it is very natural to imagine that there is a connection between the brane intersection and a negative \Renyi entropy. Naively we can derive
\bea 
S_A^{(n)}\leq S_A^{(1)}=\frac{1}{4G}\frac{x_2-x_1}{z_H},
\eea 
where indeed implies that $S_A^{(n)}$ becomes negative when $x_2<x_1$ \ie two branes intersect.

\subsubsection{Brane intersection}
\label{btzinter}
To study the brane intersection in BTZ geometries in detail let us consider the extremal case: both of the two EOW branes have negative brane tension and their profiles are
\bea 
Q_1: z_1(x)=\frac{z_H}{\lambda_1}\sinh \frac{x}{z_H},\quad Q_2: z_2(x)=\frac{z_H}{\lambda_2}\sinh \frac{x_0-x}{z_H},\quad x_0>0.
\eea 
We denote their intersection point by $(x_*,z_*)$ which is determined from
\bea 
&&e^{\frac{x_*}{z_H}}\equiv y_*=\sqrt{\frac{\a \lambda_1+\lambda_2}{\a^{-1}\lambda_1+\lambda_2}},\quad z_*= z_H\frac{1-y^2_*}{2\lambda_1 y_*}\equiv z_H \gamma,\quad \alpha\equiv e^{\frac{x_0}{z_H}}>1,
\eea 
 and their ends on the horizon by $x_1$ and $x_2$, respectively. When $\gamma<1$ the intersection point is outside of the horizon. The intersection angle $\theta_0$ can be computed as
 \bea \label{angle}
 \cos(\pi-\theta_0)=\vec{n}_1\cdot \vec{n}_2&=&\frac{2\alpha \lambda_1\lambda_2-\a^2-1}{2\a \sqrt{(1+\lambda_1^2)(1+\lambda_1^2)}}\\
 &=&\frac{(1-\gamma^2)\lambda_1\lambda_2-\sqrt{(1+\gamma^2\lambda_1^2)(1+\gamma^2\lambda_1^2)}}{ \sqrt{(1+\lambda_1^2)(1+\lambda_1^2)}},
 \eea 
 where $\vec{n}$ is the normalized normal vector of the EOW brane. If we fix the brane tensions and require $\gamma<1$ the range of the intersection angle is given by
 \bea 
 1>\cos(\theta_0)>\frac{1-\lambda_1\lambda_2}{\sqrt{(1+\lambda_1^2)(1+\lambda_1^2)}}.
 \eea  
As shown in Fig.\ref{inter} there are two candidates of the bulk dual of the BCFT on $[0,x_0]$. Let us examine them one by one.

\subsubsection*{Region I}
\begin{figure}[h!]
\centering
        \includegraphics[scale=0.4]{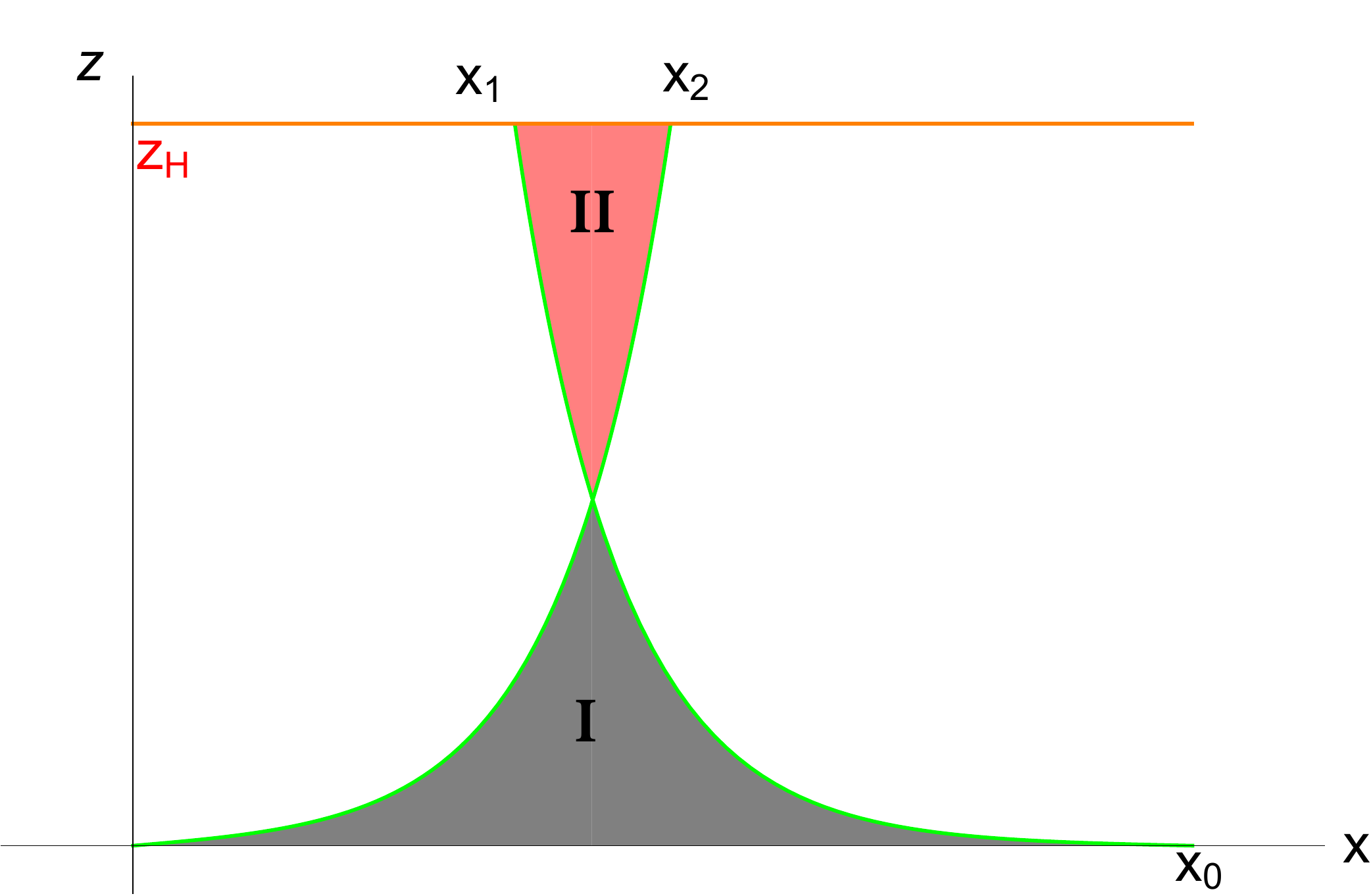}
  \caption{Two EOW branes intersect outside of the BTZ horizon.}\label{inter}
\end{figure}

Motivated by previous results \cite{Miyaji:2022cma,Biswas:2022xfw,Kusuki:2022ozk} involved with brane intersection one may think that the region I in Fig.\ref{inter} 
is the bulk dual since it is bounded by the EOW branes and the interval $[0,x_0]$. To verify this let us compute the on-shell action in the region I. The bulk contribution is 
\bea 
I_{\text{bulk}}&=&-\frac{1}{16\pi G}\int (R-2\Lambda)\sqrt{g} \\
&=&\frac{\beta}{4\pi G}\(\int_\epsilon^{z_1}\frac{1}{z^3}dz\int_0^{x_*}dx+\int_\epsilon^{z_2}\frac{1}{z^3}dz\int_{x_*}^{x_0}dx \)\\
&=&\frac{\beta}{8\pi G}\(\frac{x_0}{\epsilon^2}-\int_0^{x_*}\frac{1}{z_1^2}dx-\int^{x_0}_{x_*}\frac{1}{z_2^2}dx\)
\eea 
and the boundary contribution is 
\bea 
I_{\text{bdy}}=-\frac{1}{8\pi G}\int \sqrt{h}(K-T)=\frac{\beta}{8\pi G}\(\int_0^{x_*}\frac{1}{z_1^2}dx+\int^{x_0}_{x_*}\frac{1}{z_2^2}dx\).
\eea 
By adding the same counter term \eqref{ct} we arrive at the regularized on-shell action
\bea 
I_{\text{on-shell,I}}=\frac{\beta x_0}{16\pi Gz_H^2}=\frac{c }{12 }\frac{2\pi}{\beta }W.
\eea 
Clearly, this bulk geometry is not the desired one. From the relation between on-shell action and ADM mass $I=2\pi M_{\text{ADM}}$ \cite{Wald:1993nt}, it implies the bulk geometry is possible to dual to a state with a very high energy $E=M_{\text{ADM}}=\frac{c}{12}W/\beta$.

\subsubsection*{Region I $\cup$ II}
Another obvious guess is that the bulk dual is Region I $\cup$ II. However, according to the rule (the normal vectors of the boundaries should point outward) of AdS/BCFT, the tension of the two EOW branes bounding region II should be positive. It is a little suspicious because it means that after crossing the sign of the brane tension of these two EOW branes gets flipped! Indeed we can show that this guess is not correct by computing the on-shell action. The bulk contribution in region II is 
\bea 
\tilde{I}_{\text{bulk}}&=&\frac{\beta}{4\pi G}\(\int_{z_2}^{z_H}\frac{1}{z^3}dz \int_{x_2}^{x_*}dx+\int_{z_1}^{z_H}\frac{1}{z^3}dz \int^{x_1}_{x_*}dx\)\\
&=&\frac{\beta}{8\pi G}\(-\frac{x_1-x_2}{z_H^2}+\int_{x_2}^{x_*}\frac{1}{z_2^2}dx+\int^{x_1}_{x_*}\frac{1}{z_1^2}dx\),
\eea 
and the boundary contribution is 
\bea 
\tilde{I}_{\text{bdy}}=-\frac{\beta}{8\pi G}\(\int_{x_2}^{x_*}\frac{1}{z_2^2}dx+\int^{x_1}_{x_*}\frac{1}{z_1^2}dx\).
\eea 
Then we find
\bea 
I_{\text{on-shell}}(\beta)&=&I_{\text{on-shell,I}}+\tilde{I}_{\text{bulk}}+\tilde{I}_{\text{bdy}}=\frac{1}{4G}\(\frac{x_0}{2z_H}+\frac{x_2-x_1}{z_H}\)\\
&=&\frac{1}{4G}\(\frac{3x_0}{2z_H}+\frac{x_0-x_2}{z_H}-\frac{x_1}{z_H}\)\\
&=&\frac{c}{12}\frac{2\pi}{\beta}3W-\frac{c}{6}\sinh ^{-1}(\lambda_2 ) -\frac{c}{6}\sinh ^{-1}(\lambda_1 ) 
\eea 
which is different from the expected result \eqref{expect}. However, we find that if we subtract the on-shell action of Region II we can obtain the correct result 
\bea 
I_{\text{on-shell}}(\beta)=I_{\text{on-shell,I}}-\tilde{I}_{\text{bulk}}-\tilde{I}_{\text{bdy}}=-\frac{c}{12}\frac{2\pi}{\beta}W+\frac{c}{6}\sinh ^{-1}(\lambda_2 ) +\frac{c}{6}\sinh ^{-1}(\lambda_1 ),
\eea 
and it leads to the desired \Renyi entropy using \eqref{entropyc}. Therefore the result suggests that we should not flip the sign of the brane tension and propose that when the normal vectors of the boundaries point inward the orientation of this bulk region should be inverted. In other words, the correct bulk dual is Region I$\cup {\text{II}^{-1}}$. One attitude to this bizarre result is that the bulk solution with brane intersection is non-physical just like the bulk solution with brane self-intersection \cite{Kusuki:2022ozk,Cooper:2018cmb,Geng:2021iyq,Bianchi:2022ulu,Kawamoto:2022etl}. It implies that the simple gravity model \eqref{gravity} is not adequate when the brane tension is large. From the field theory side, the brane self-intersection is excluded by a bootstrap analysis \cite{Kusuki:2022ozk}. Naively our results suggest that the brane intersection can be excluded by the positivity of \Renyi entropy. Note that \Renyi entropy of the interval at the end of a semi-infinite line is related to a one-point function of the BCFT which is a crucial input of the bootstrap analysis. Then it is possible to understand the brane non-intersection also from the bootstrap analysis but it is beyond the scope of this paper.

\subsubsection{From Dong's formula}
In the end, let us show how this result can also be obtained from Dong's formula. It is proposed by Dong Xi, the holographic \Renyi entropy of a subsystem $A$ can be computed from a refined version of \Renyi entropy defined as
\bea \label{reen}
\tilde{S}_n^A=n^2\partial_n\(\frac{n-1}{n}S_n^A\),
\eea 
which is dual to the area of a bulk codimension-2 cosmic brane homologous to the subregion $A$:
\bea 
\tilde{S}_n^A=\frac{\text{Area}(\text{Cosmic Brane}_n)}{4G}.
\eea 
The cosmic brane has a tension $(n-1)/(4n G)$ so it back reacts on the bulk geometry by creating a  conical defect with an opening angle $2\pi/n$. So the goal is to find such a conical geometry with the boundary to be our cylinder. Starting from the BTZ back hole metric we can make the replacement $h(z)\rightarrow h_n(z)=1- z^2/(n^2z_H^2)$ and fix the period $\tau\sim \tau+2\pi z_H$ then the resulting geometry will have a required conical singularity at $z=nz_H$. This geometry is exactly the conical geometry that we are looking for. Thus the refined \Renyi entropy is
\bea \label{enth}
\tilde{S}=\frac{1}{4G}\int_{z=nz_H}\sqrt{h}=\frac{1}{4G}\frac{x_2-x_1}{n z_H}=\frac{1}{4G}\(\frac{x_0}{nz_H }+\frac{x_2-x_0}{nz_H}+\frac{-x_1}{nz_H }\)=\frac{c}{6}\frac{W}{n}+g_A+g_B,
\eea 
which exactly leads to \eqref{renyi2}. It is interesting to see that Dong's formula not only gives the leading log terms but also the boundary entropy terms \footnote{as far as we notice this is the first time to reproduce the boundary entropy using Dong's formula}. When $x_2<x_1$, \ie when the branes intersect we find that the refined \Renyi entropy becomes negative but it does not mean the area of the cosmic brane is negative. Our interpretation is that the area is still positive but the orientation is inverted! Next, we consider another possible bulk solution for reproducing the \Renyi entropy.

\subsection{Annulus boundary}
In the field theory analysis described in section \ref{annulus}, we have shown  when the original spacetime $\mathcal{M}_1$ has a geometry of annulus the replicated spacetime $\mathcal{M}_n$ will have a conical singularity which should be regularized. We will show that bulk dual of $\mathcal{M}_n$  can be constructed from the \Banados map \cite{Banados:1998gg,Roberts:2012aq}.

\subsubsection{The original partition function $Z_1$}
When the boundary BCFT is defined on an annulus (or a disk), the bulk geometry can be \Poincare AdS with metric 
\bea \label{p1}
ds^2=\frac{r^2d\tau^2+dr^2+dz^2}{z^2},
\eea 
and the EOW brane profile\footnote{When the normal vector of the brane points towards the center the brane tension is negative for the choice $(+)$ and positive for the choice $(-)$ .} \cite{Fujita:2011fp}
\bea \label{round}
r^2+(z\pm r_D\lambda)^2=r_D^2(1+\lambda^2),
\eea 
where $\lambda$ is still defined as \eqref{lambda} and $r_D$ is a free parameter. 
It is helpful to consider the case of a disk with radius $r_D$ first. The bulk dual geometry is a part of the round sphere described by \eqref{round}. Then the on-shell action in this region can be easily computed as the following. The bulk contribution is 
\bea 
I_{\text{bulk}}=\frac{4\times 2\pi}{16\pi G}\int_{\epsilon}^{z_M}\frac{dz}{z^3}\int_0^{r(z)}r dr=\frac{1}{4G}\(\frac{r_D(-r_D\pm 4\lambda z)}{2z^2}-\log z\) \Big|_{\epsilon}^{z_M},
\eea 
the boundary contribution 
\bea 
I_{\text{bdy}}&=&-\frac{1}{4 G}\int_\epsilon^{z_M}\frac{r_D \sqrt{1+\lambda^2}}{z^2} dz(\mp \frac{\lambda}{\sqrt{1+\lambda^2}}) =\mp \frac{r_D\lambda}{4 G}\(\frac{1}{z_M}-\frac{1}{\epsilon}\),
\eea 
and the counter term is 
\bea 
I_{\text{ct}}=-\frac{1}{8\pi G}\int_{z=\epsilon} \sqrt{h} dr=\frac{1}{4G}\(\frac{r_D^2}{2\epsilon^2}\mp\frac{\lambda r_D}{\epsilon}-\frac{1}{2}\),
\eea 
thus the final result is \cite{Takayanagi:2011zk,Fujita:2011fp}
\bea \label{field}
I_{\text{on-shell}}=\frac{1}{4G}(\log\frac{\epsilon}{r_D}\pm \sinh^{-1}\lambda).
\eea 
The logarithm term is still divergent but expected because it is responsible for the Weyl anomaly. 
To obtain a finite result we can consider an annulus instead of a disk on the boundary. 
Let the radius of the annulus be $r_+$ and $r_-$ then there are still two situations: the two branes intersect or not in the bulk. When they do not intersect the on-shell action of the corresponding bulk region is simply given by 
\bea \label{z1}
I_{\text{on-shell}}(r_+)-I_{\text{on-shell}}(r_-)=\frac{1}{4G}(\log\frac{r_-}{r_+}\pm \sinh^{-1}\lambda_1\mp \sinh^{-1}\lambda_2),
\eea 
which matches the field theory result \eqref{fz1}.
\subsubsection{Brane intersection}
\begin{figure}[h]
\centering
	\subfloat[]{
	\centering
	\includegraphics[scale=0.25]{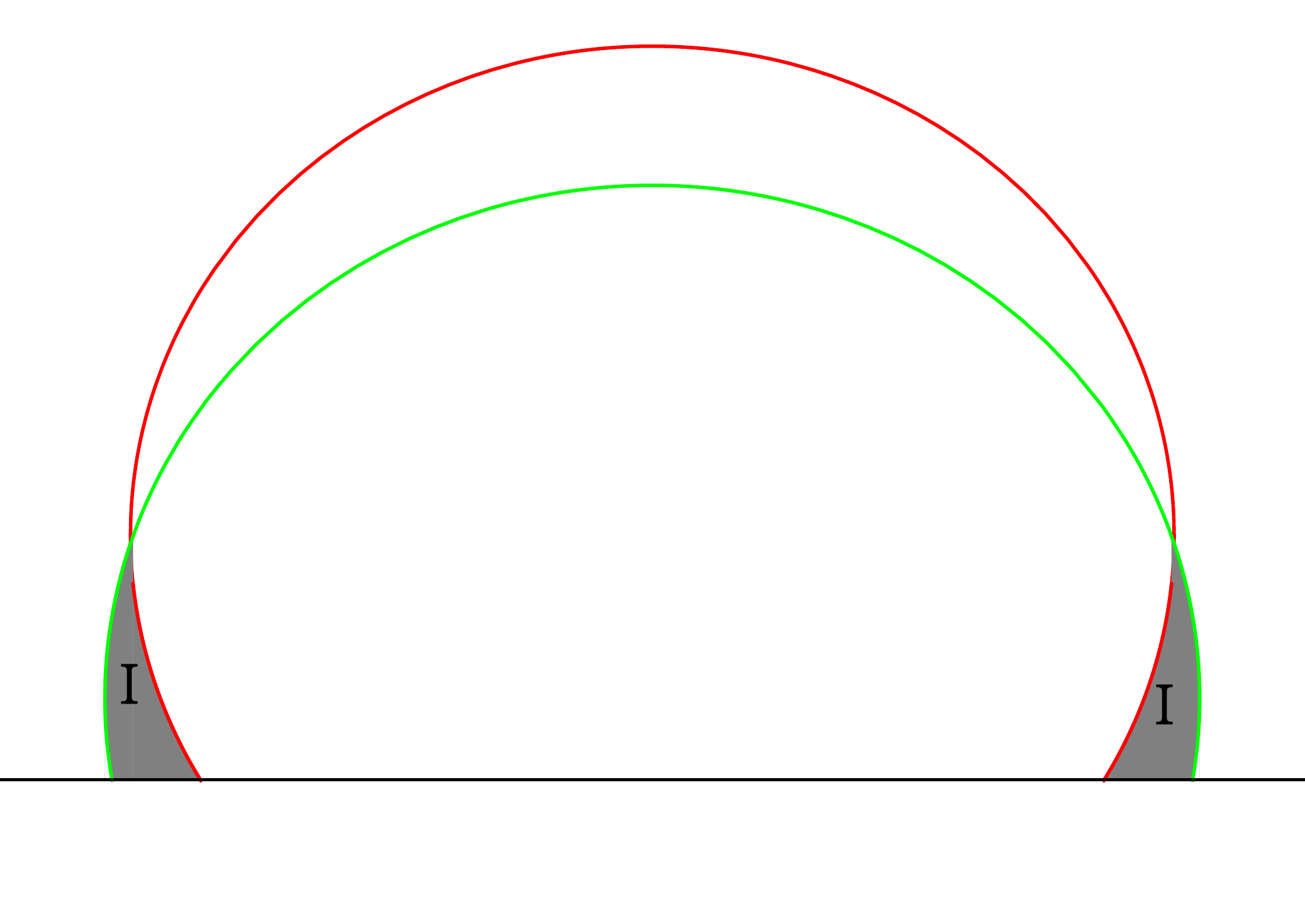}}
        \hfill
    \subfloat[]{
    \centering
    \includegraphics[scale=0.25]{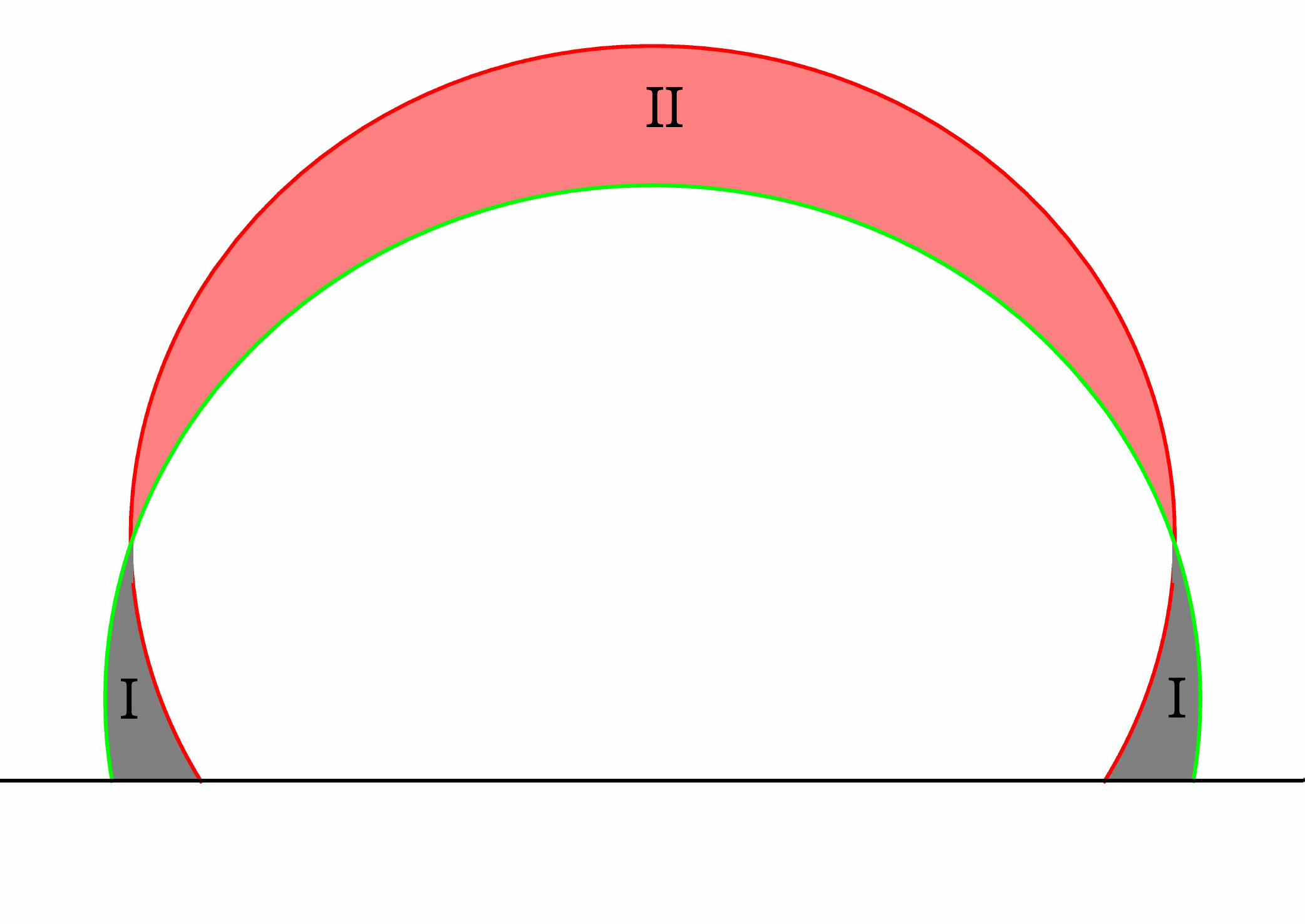}}
  \caption{$(a)$: The shaded region is the naive guess of the bulk dual geometry. $(b)$: we propose that the region I$\cup$II  is the correct bulk dual.}\label{dis}
\end{figure}

As shown in Fig.\ref{dis}, when we tune the brane tension two EOW branes are possible to intersect. 
To be concrete let us consider two EOW branes
\bea 
&& Q_1: r^2+(z+r_-\lambda_1)^2=r_-^2(1+\lambda_1^2),\\
&& Q_2: r^2+(z+r_+\lambda_2)^2=r_+^2(1+\lambda_2^2),
\eea 
where we assume that $0<r_-<r_+$. The two branes will intersect when 
\bea 
&& r_-(\sqrt{1+\lambda_1^2}-\lambda_1)\geq r_+(\sqrt{1+\lambda_2^2}-\lambda_2),\quad \rightarrow \\
&&1<\frac{r_+}{r_-}\equiv a\leq \frac{\sqrt{1+\lambda_2^2}+\lambda_2}{\sqrt{1+\lambda_1^2}+\lambda_1},
\eea 
at the position $(r_*,z_*)$
\bea 
z_*=\frac{r_+^2-r_-^2}{2(r_+\lambda_2-r_-\lambda_1)},\quad r_*^2+(z_*-r_-\lambda_1)^2=r_-^2(1+\lambda_1^2).
\eea 
Similar to \eqref{angle} the intersection angle is given by
\bea 
\cos(\pi-\theta_0)=\frac{1+a^2+2a\lambda_1\lambda_2}{2a\sqrt{(1+\lambda_1^2)(1+\lambda_2^2)}}.
\eea 
Naively we guess that the bulk dual is region I in Fig.\ref{dis} however it is straightforward to show that the on-shell action in region I vanishes thus we have to also consider region II. Note that in region I the normal vector of $Q_1(Q_2)$ should point inward (outward) so if we preserve the sign of brane tensions we should change the orientation of region II. Thus 
\iffalse 
The bulk contribution is
\bea 
I_{\text{bulk}}&=&\frac{4\times 2\pi}{16\pi G}\int_{\epsilon}^{z_*}\frac{dz}{z^3}\int_{r_1(z)}^{r_2(z)}r dr\\
&=&\frac{1}{4G}\(\frac{r_+(-r_+- 4\lambda_2 z)}{2z^2}-\frac{r_-(-r_-- 4\lambda_1 z)}{2z^2}\) \Big|_{\epsilon}^{z_*}
\eea 
and the boundary contribution is
\bea 
I_{\text{bdy}}=-\frac{r_-\lambda_1}{4G}(\frac{1}{z_*}-\frac{1}{\epsilon})+\frac{r_+\lambda_2}{4G}(\frac{1}{z_*}-\frac{1}{\epsilon}).
\eea 
Finally, we  add a counter term \eqref{ct} to obtain the on-shell action
\bea 
I_{\text{on-shell}}=0.
\eea 
Apparently, this is not the correct on-shell action to compute the \Renyi entropy. However, if we also include region II as shown in Fig.\eqref{dis} we can get the correct result. 
\fi 
the on-shell action of region II$^{-1}$ is computed as
\bea 
\tilde{I}&=&\frac{1}{4G}\(\frac{r_+^2}{2z_*^2}+\frac{\lambda_2 r_+}{z_*}-\frac{1}{2}+\log \frac{z_*}{r_+}-\sinh^{-1}\lambda_2\)\\
&-&\frac{1}{4G}\(\frac{r_-^2}{2z_*^2}+\frac{\lambda_2 r_-}{z_*}-\frac{1}{2}+\log \frac{z_*}{r_-}-\sinh^{-1}\lambda_1\)\\
&=&\frac{1}{4G}\(\log\frac{r_-}{r_+}+\sinh^{-1}\lambda_1-\sinh^{-1}\lambda_2\),
\eea 
which exactly gives the desired on-shell action $Z_1$ \eqref{z1}. 

\subsubsection{The replica partition function $Z_n$}
Now we want to find the bulk geometry for computing $Z_n$. Note that the 2d CFT is defined on conical manifold $\mathcal{M}_n$ with the metric
\bea 
ds^2_{\mathcal{M}_n}=dr^2+n^2r^2 d\theta^2=d\zeta d\bar{\zeta},\quad \zeta=r e^{\im n \phi}.
\eea 
Introducing $\xi=\zeta^{1/n}$ then the metric can be written as
\bea \label{cover}
ds^2_{\mathcal{M}_n}=n^2|\xi|^{n-1}d\xi d\bar{\xi}=n^2|\xi|^{n-1}ds^2_{\hat{\mathcal{M}}_n}.
\eea 
Thus the 2d CFT defined on $\mathcal{M}_n$ is conformally related to the 2d CFT defined on $\hat{\mathcal{M}}_n$. Considering the vacuum state of the latter theory, the 3d bulk dual is the \Poincare AdS$_3$ with metric 
\bea 
ds^2=\frac{dw d\bar{w }+dz^2}{z^2}.
\eea 
The conformal transformation $\w=\xi^{n}\equiv p(\xi),\quad \bar{w}=\bar{\xi}^{n}\equiv q(\bar{\xi })$ can be lifted to a 3d diffeomorphism known as the \Banados map\footnote{It is a little different from the ones \cite{Roberts:2012aq} because we work in the Euclidean signature.}
\bea \label{banado}
&&w =p(\xi )-\frac{2\eta^2(p')^2 q''}{4p'q'+\eta^2p''q''},\\
&&\bar{w} = q(\bar{\xi })-\frac{2\eta^2(q')^2 p''}{4p'q'+\eta^2p''q''},\\
&&z = \frac{4\eta(p'q')^{3/2}}{4p'q'+\eta^2p''q''}.
\eea 
The corresponding bulk geometry has the metric
\bea 
ds^2=\frac{d\eta^2}{\eta^2}+T_+(\xi )d\xi^2+T_- d\bar{\xi }^2+\(\frac{1}{\eta^2}+\eta^2 T_+T_-\)d\xi  d\bar{\xi },
\eea 
where
\bea 
T_+(\zeta)=\frac{3(p'')^2-2p'p'''}{4{p'}^2},\quad T_-(\bar{\zeta})=\frac{3(q'')^2-2q'q'''}{4{q'}^2}.
\eea 
In our case, the map and $T_\pm $ are
\bea 
&& w=\frac{\xi^{n} \left(4 {\bar{\xi  }} \xi  +\eta ^2 \left(n^2-1\right)\right)}{4 \bar{\xi  } \xi   +\eta ^2 (n-1)^2},\quad \bar{w}=\frac{\bar{\xi}^{n} \left(4 {\bar{\xi  }} \xi  +\eta ^2 \left(n^2-1\right)\right)}{4 \bar{\xi  } \xi   +\eta ^2 (n-1)^2},\\
&&z=\frac{4 \eta  n {\bar{\xi }}^{\frac{n+1}{2 }} \xi  ^{\frac{n+1}{2 }}}{4 {\bar{\xi }} \xi   +\eta ^2 (n-1)^2},\quad T_+=\frac{n^2-1}{4\xi^2},\quad T_-=\frac{n^2-1}{4\bar{\xi }^2}.
\eea 
The metric can be written in the form
\bea \label{replicametric}
ds^2=\frac{d\eta^2+f_+(\eta,\rho)d\rho^2+f_-(\eta,\rho)\rho^2 d\theta^2}{\eta^2},
\eea 
with
\bea 
\xi \equiv \rho e^{\im \theta},\quad f_\pm=\(\frac{(n^2-1)\eta^2\pm 4\rho^2}{4 \rho^2}\)^2.
\eea 
At the position $\eta_c=\frac{2\rho}{\sqrt{n^2-1}}$ the metric degenerates so we should treat it as the center (or horizon) of the bulk space \cite{Skenderis:1999nb} (see also \cite{Hung:2011nu}). First, it is useful to compute the on-shell action in the whole spacetime with an IR cut-off $\rho=\Lambda_\rho$:
\bea 
I_{\text{bulk}}&=&-\frac{1}{16\pi G}\int \sqrt{g}(R-\Lambda)=\frac{1}{2G}\int_{\epsilon}^{\eta_c}\int d\rho \(\frac{\rho }{\eta ^3}-\frac{\eta  \left(n^2-1\right)^2}{16  \rho ^3}\)\\
&=&\frac{1}{4G}\int^{\Lambda_\rho}_{\frac{\sqrt{n^2-1}}{2}\epsilon} d\rho\(\frac{\rho}{\epsilon^2}+\frac{\epsilon^2(n^2-1)^2}{16\rho^3}-\frac{n^2-1}{2\rho}\),\\
I_{\text{GH}}&=&-\frac{1}{8\pi G}\int_{\eta=\epsilon} \sqrt{\gamma_\epsilon}K_\epsilon\\
&=&-\frac{1}{4 G}\int^{\Lambda_\rho}_{\frac{\sqrt{n^2-1}}{2}\epsilon} d\rho\(\frac{2\rho}{\epsilon^2}+\frac{\epsilon^2(n^2-1)^2}{8\rho^3}\),\\
I_{ct}&=&\frac{1}{4G}\int_{\eta=\epsilon} \sqrt{\gamma_\epsilon}=\frac{1}{4G}\int^{\Lambda_\rho}_{\frac{\sqrt{n^2-1}}{2}\epsilon} d\rho \(\frac{\rho}{\epsilon^2}-\frac{\epsilon^2(n^2-1)^2}{16\rho^3}\),\\
I_{\text{on-shell,}\Lambda_\rho}&=&I_{\text{bulk}}+I_{\text{GH}}+I_{ct}\\
&=&\frac{c}{12}{(1-n^2)}\(\log\frac{\Lambda_\rho}{\epsilon}+\frac{1}{2}(\log\frac{4}{n^2-1}+1)\).\label{IR}
\eea

\begin{figure}[h!]
\centering
        \includegraphics[scale=0.3]{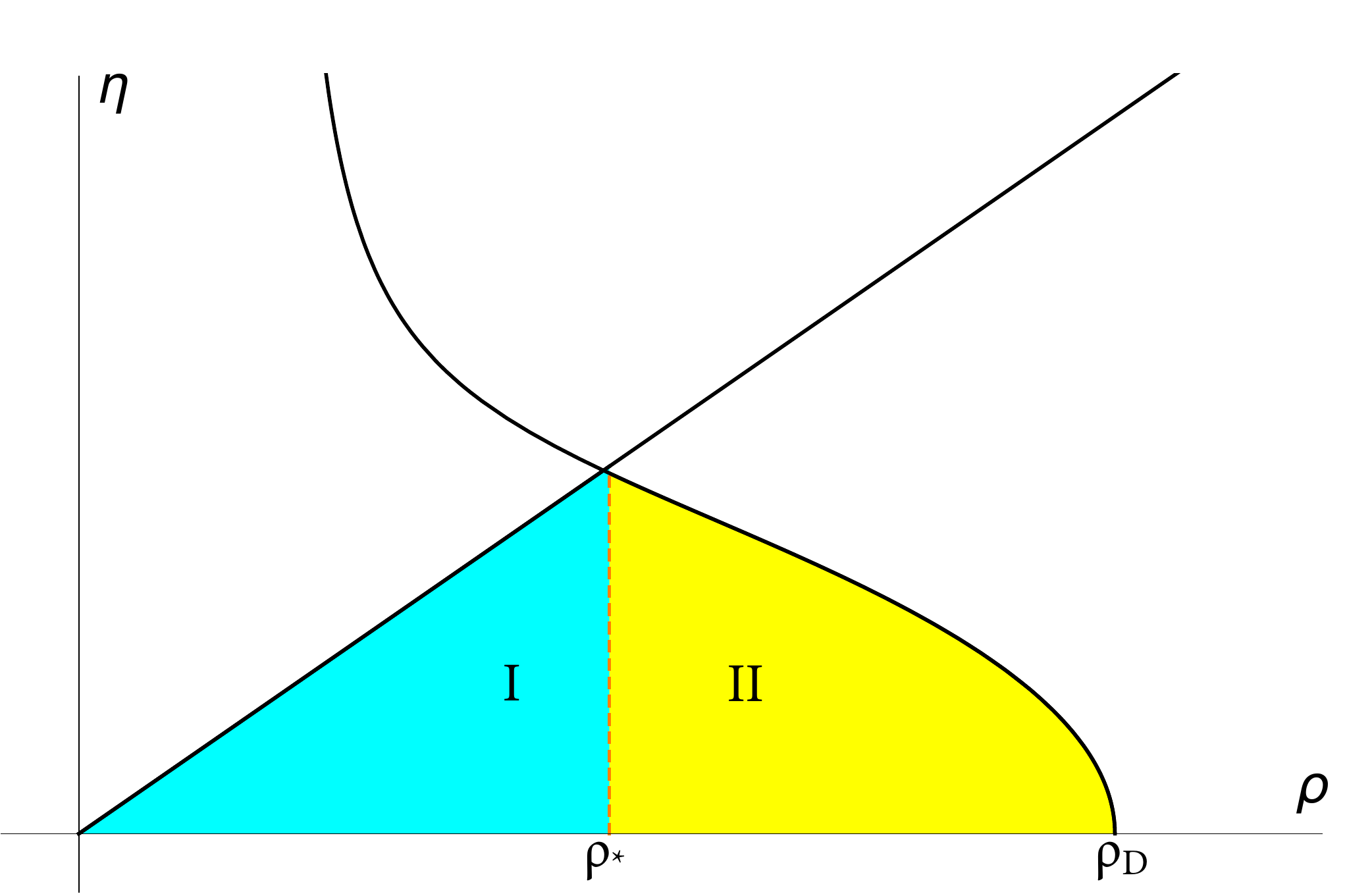}
  \caption{The bulk region corresponding to a disk region in the boundary.}\label{eow1}
\end{figure}

Next, we replace the IR cut-off $\rho=\Lambda_\rho$ with an EOW brane as shown in Fig.\ref{eow1}.
Using the map \eqref{banado} the profile of EOW branes becomes
\bea \label{replicaeow}
\((n+1)^2\rho^{{2}{n}}-(n-1)^2\rho_D^{{2}{n}}\)\eta^2-8n\lambda \rho_D^{{n}}\rho^{1+{n}}\eta+4\rho^2(\rho^{{2}{n}}-\rho_D^{{2}{n}})=0,
\eea 
where $\rho_D=r_D^{\frac{1}{n}}$. For simplicity, let us first choose $\lambda=0$. Then the profile is
\bea 
\eta_E &=&{2\rho}\sqrt{\frac{\rho_D^{2n}-\rho^{2n}}{(n+1)^2\rho^{2n}-(n-1)^2\rho_D^{2n}}}\\
&=&\frac{2\rho}{\sqrt{n^2-1}}\sqrt{\frac{{n-1}}{n+1}}\sqrt{\frac{\rho_D^{2n}-\rho^{2n}}{\rho^{2n}-(\frac{n-1}{n+1})^2\rho_D^{2n}}},
\eea 
which implies that $\rho$ has to be in the range
\bea 
\(\frac{n-1}{n+1}\)^{\frac{1}{n}}\rho_D^{}<\rho<\rho_D.
\eea 
For convenience, we divide the dual bulk spacetime into 2 regions and the on-shell action in region II of Fig.\ref{eow1}
 can be easily computed as
\bea
I_{\text{on-shell},II}&=&I_{\text{bulk}}+I_{GH}+I_{ct}\\
&=&-\frac{1}{16\pi G}\int \sqrt{g}(R-\Lambda)=\frac{1}{2G}\int_{\epsilon}^{\eta_E}\int d\rho \(\frac{\rho }{\eta ^3}-\frac{\eta  \left(n^2-1\right)^2}{16  \rho ^3}\)\\
&=&\frac{1}{4G}\int^{\rho_D}_{\rho*}\(-\frac{\epsilon^2(n^2-1)^2}{8\rho^3}-\frac{\rho }{\eta_E ^3}-\frac{\eta_E ^2 \left(n^2-1\right)^2}{16  \rho ^3}\)\label{ep}\\
&=&-\frac{1}{4G}\int^{\rho_D}_{\rho_*}\(\frac{\rho }{\eta_E ^3}+\frac{\eta_E^2  \left(n^2-1\right)^2}{16 \rho ^3}\)\\
&=&\frac{1}{4G}\(\frac{(n^2+1)}{2}\log\rho+\log\frac{\eta_E}{2\rho}\) \Big|_{\rho_*}^{\rho_D},\quad \rho_*=\rho_D \(\frac{n-1}{n+1}\)^{\frac{1}{2n}}\\
&=&-\frac{c}{6}\(\log\frac{\rho_D}{\epsilon}+\log\frac{2}{\sqrt{n^2-1}}+\frac{n^2+1}{4n}\log\frac{n-1}{n+1}\).
\eea 
Since $\rho$ will not approach $0$ so we can safely drop the first term in \eqref{ep}. The on-shell action in region I is given by \eqref{IR} with the replacement $\Lambda_\rho=\rho_*$:
\bea 
I_{\text{on-shell},I}=\frac{c}{12}{(1-n^2)}\(\log\frac{\rho_*}{\epsilon}+\frac{1}{2}(\log\frac{4}{n^2-1}+1)\)
\eea 
thus the final result is
\bea 
I_{\text{on-shell},I}+I_{\text{on-shell},II}\,=\hspace{-0.5cm}&&-\frac{c}{12}(n^2+1)\(\log\frac{\rho_D}{\epsilon}+\log\frac{2}{\sqrt{n^2-1}}\)\\
&&-\frac{c}{12n}\log\frac{n-1}{n+1}+\frac{c}{24}(1-n^2).
\eea 
The leading term matches the field theory calculations of the disk partition function. For the annulus, we get precisely the same result as \eqref{weyln}
\bea 
I_{\text{on-shell}}(\mathcal{M}_n)=I_{\text{on-shell}}(\rho_{\text{max}})-I_{\text{on-shell}}(\rho_{\text{min}})=-\frac{c}{12}(n^2+1)\log\frac{\rho_{\text{max}}}{\rho_{\text{min}}},
\eea 
where non-universal subleading terms are canceled. Because we have set $\lambda=0$ thus the corresponding boundary entropy also vanishes. The analytic result of the on-shell action is hard to obtain for non-vanishing $\lambda$ because of the complicity of the expression of the EOW profile. So here we will make a perturbative analysis with respect to $\lambda$ and the goal is to reproduce the boundary entropy terms. 
 
\subsection*{$\lambda \neq 0$: perturbative results}
For non-vanishing $\lambda$, we can still separate the spacetime into two parts and compute the on-shell action separately according to the line
\bea 
\rho_*=\rho_D \(\sqrt{\frac{n-1}{n+1}}(\lambda+\sqrt{1+\lambda^2})\)^{\frac{1}{n}}.
\eea 
The on-shell action in the region $\text{I}$ is still 
\bea \label{onshell1l}
I_{\text{on-shell},I}=\frac{c}{12}{(1-n^2)}\(\log\frac{\rho_*}{\epsilon}+\frac{1}{2}(\log\frac{4}{n^2-1}+1)\).
\eea 
The bulk contribution to the on-shell action in the region $\text{II}$ is
\bea 
I_{\text{bulk},II}=-\frac{1}{4G}\int^{\rho_D}_{\rho_*}\(\frac{\rho }{\eta_E ^2}+\frac{\eta_E^2  \left(n^2-1\right)^2}{16 \rho ^3}\),
\eea 
where $\eta_E$ is the profile of the EOW
\bea 
\eta_E=-\frac{2 \rho  \left(\sqrt{2 \left(2 \lambda ^2 n^2+n^2+1\right) \rho ^{2 n} \rho _D^{2 n}-(n-1)^2 \rho _D^{4 n}-(n+1)^2 \rho ^{4 n}}+2 \lambda  n \rho ^n \rho _D^n\right)}{(n-1)^2 \rho _D^{2 n}-(n+1)^2 \rho ^{2 n}}.
\eea 
In order to evaluate the boundary contribution we first work out the determinant of the induced metric 
\bea 
\sqrt{\gamma}=\frac{\left(\left(n^2-1\right) \eta_E(\rho )^2-4 \rho ^2\right) \sqrt{16 \rho ^4 \eta_E'(\rho )^2+\left(\left(n^2-1\right) \eta_E(\rho )^2+4 \rho ^2\right)^2}}{16 \rho ^3 \eta_E(\rho )^2}. 
\eea 
Then we find that leading order correction is\footnote{Note that we should regularize $\rho_D \rightarrow \rho_D-\frac{\rho_D \lambda^2}{2}-\frac{\epsilon^2}{2\rho_D}+\epsilon \lambda$.}
\bea 
I_{\text{bulk},II}^{(1)}=-\frac{1}{4G}\int_{\rho_*}^{\rho_D} \frac{4 n^2 \rho ^{n-1} \rho _D^n \left((n-1)^2 \rho _D^{4 n}-(n+1)^2 \rho ^{4 n}\right)}{\left(-\left(\left(\rho _D^{2 n}-\rho ^{2 n}\right) \left((n-1)^2 \rho _D^{2 n}-(n+1)^2 \rho ^{2 n}\right)\right)\right){}^{3/2}}
\eea 
and 
\bea 
I_{\text{EOW},II}^{(1)}=\frac{1}{8G}\int_{\rho_*}^{\rho_D} \frac{4 n^2 \rho ^{n-1} \rho _D^n \left((n-1)^2 \rho _D^{4 n}-(n+1)^2 \rho ^{4 n}\right)}{\left(-\left(\left(\rho _D^{2 n}-\rho ^{2 n}\right) \left((n-1)^2 \rho _D^{2 n}-(n+1)^2 \rho ^{2 n}\right)\right)\right){}^{3/2}}
\eea 
thus
\bea 
&&I_{\text{bulk},II}^{(1)}+I_{\text{EOW},II}^{(1)}\\
&&=-\frac{1}{8G}\int_{\rho_*}^{\rho_D} \frac{4 n^2 \rho ^{n-1} \rho _D^n \left((n-1)^2 \rho _D^{4 n}-(n+1)^2 \rho ^{4 n}\right)}{\left(-\left(\left(\rho _D^{2 n}-\rho ^{2 n}\right) \left((n-1)^2 \rho _D^{2 n}-(n+1)^2 \rho ^{2 n}\right)\right)\right){}^{3/2}}\\
&&=-\frac{c}{6} \frac{n^2+1}{2n} \lambda.
\eea 
 Adding the leading term of \eqref{onshell1l}:
\bea 
I_{\text{on-shell},I}^{(1)}=\frac{c}{12}(1-n^2)\frac{\lambda}{n}
\eea 
we get
\bea 
I_{\text{on-shell},I}^{(1)}+I_{\text{on-shell},II}^{(1)}=-\frac{c}{6}n\lambda,
\eea 
as expected considering that 
\bea 
\sinh ^{-1}(\lambda)\sim \lambda+\mathcal{O}(\lambda^2).
\eea 
We also work out the subleading correction
\bea 
&&I_{\text{bulk},II}^{(3)}+I_{\text{EOW},II}^{(3)}=-\frac{c}{6} \(\frac{n^2+1}{2n} (\lambda -\frac{1}{6}\lambda^3)\),\\
&&I_{\text{on-shell},I}^{(3)}=\frac{c}{12}(1-n^2)\frac{1}{n}(\lambda-\frac{1}{6}\lambda^3),\\
&&I_{\text{on-shell},I}^{(3)}+I_{\text{on-shell},II}^{(3)}=-\frac{c}{6}n\(\lambda-\frac{1}{6}\lambda^3\),
\eea 
which also matches the expansion of the field theory result \eqref{field}. Thus we conclude that the bulk metric \eqref{replicametric} with the EOW brane profile \eqref{replicaeow} gives the correct bulk dual for computing the replica partition function $Z_n$.

\subsection*{Brane intersection}
\begin{figure}[h!]
\centering
        \includegraphics[scale=0.4]{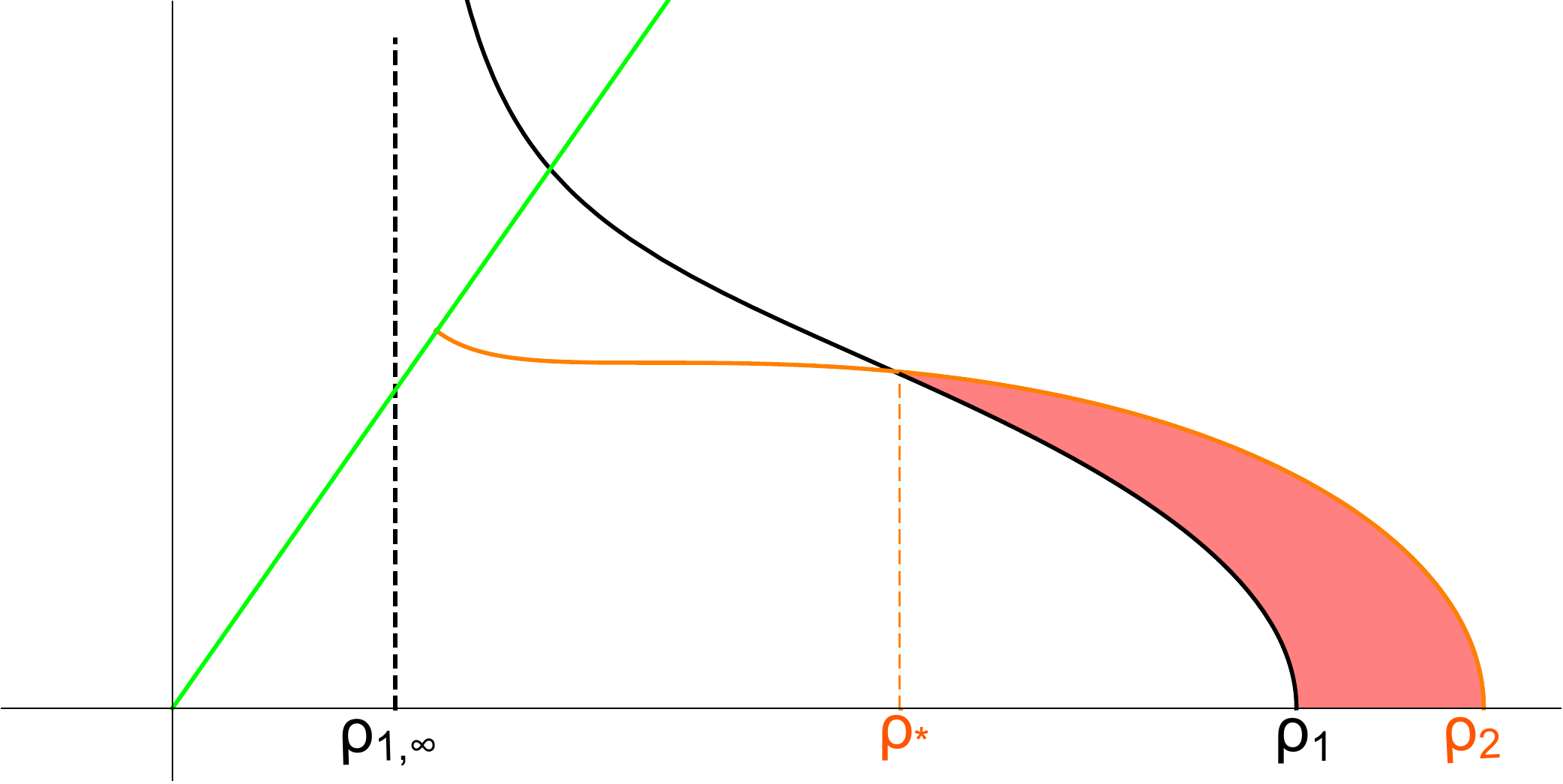}
  \caption{The bulk geometry with brane intersection corresponding to an annulus region on the boundary.}\label{conn}
\end{figure}

For the annulus boundary, the bulk geometry may also have an intersection phase as shown in Fig.\ref{conn}. It is similar to the situation in the BTZ case discussed in \ref{btzinter}. Even though the shaded region in Fig.\ref{conn} is not the proper bulk dual for computing $Z_n$ but it is still interesting to see what is its possible dual state.

For simplicity, we assume that one of the EOW branes is tensionless 
\bea 
&& Q_1: \eta={2\rho}\sqrt{\frac{\rho_1^{2n}-\rho^{2n}}{(n+1)^2\rho^{2n}-(n-1)^2\rho_1^{2n}}},\nn
&& Q_2: \eta=\frac{2 \left(\sqrt{-\rho ^2 \left(-2 \left(\left(2 \lambda ^2+1\right) n^2+1\right) \rho ^{2 n} \rho _2^{2 n}+(n+1)^2 \rho ^{4 n}+(n-1)^2 \rho _2^{4 n}\right)}-2 \lambda  n \rho ^{n+1} \rho _2^n\right)}{(n-1)^2 \rho _2^{2 n}-(n+1)^2 \rho ^{2 n}}.\nonumber  
\eea 
The intersection point is at
\bea 
\rho_s^{2n}=\frac{\rho_1^{2n}+\rho^{2n}_{1\infty}-\alpha^2+\sqrt{(\rho_1^{2n}+\rho^{2n}_{1\infty}-\alpha^2)^2-4\rho_1^{2n}\rho^{2n}_{1\infty}}}{2},
\eea 
where 
\bea 
\rho_{1\infty}=\rho_1\(\frac{n-1}{n+1}\)^{\frac{1}{n}},\quad \alpha=\frac{\rho_2^{2n}-\rho_1^{2n}}{\lambda (n+1)}.
\eea 
The on-shell action can be computed from the difference 
\bea 
I_{\text{on-shell}}&=&I_{\text{on-shell},Q_2}([\rho_s,\rho_2])-I_{\text{on-shell},Q_1}([\rho_s,\rho_1])\\
&=& \frac{c}{12}(n^2-1)\log\frac{\rho_2}{\rho_1}+\mathcal{O}(\lambda^2),
\eea 
which is totally different from the expected result \eqref{weyln}.
According to the identification $I=2\pi M_{\text{ADM}}$, it suggests this bulk geometry is dual to a state with energy  with energy
\bea 
M_{\text{ADM}}=\frac{c}{12}(n^2-1).
\eea 
\subsubsection{From Dong's formula}
In the end, let us argue how to use Dong's formula to directly obtain the result \footnote{The argument is similar in spirit if not in detail to one in \cite{Kusuki:2022ozk}}. The replica manifold $\mathcal{M}_n$ has a conical singularity so we first introduce a uniform cover $\hat{\mathcal{M}}_n$ defined in \eqref{cover} to regularize it.  Now we can put a flat metric on the uniform cover thus the bulk dual $\hat{\mathcal{B}}_n$ is simply the \Poincare metric. Then the required conical geometry can be obtained by taking the $\mathbb{Z}_n$ quotient of $\hat{\mathcal{B}}_n$. 
Thus the cosmic brane which is the stabilizer of $\mathbb{Z}_n$ quotient just starts from the center of the small EOW brane and ends at the center of the big EOW brane so that refined \Renyi entropy which is given by the length of this cosmic brane which is exactly \eqref{weyln}
\bea
\frac{1}{4G}\int^{r^{1/n}_{+}(\sqrt{1+\lambda_2^2}+\lambda_2)}_{r^{1/n}_{-}(\sqrt{1+\lambda_1^2}+\lambda_1)} \frac{dz }{z}=\frac{c}{6 }\(\frac{1}{n}\log\frac{r_+}{r_-}+\sinh ^{-1}(\lambda_2 )-\sinh ^{-1}(\lambda_1 )\).
\eea
From Dong's formula it is easier to see when $r^{1/n}_{+}(\sqrt{1+\lambda_2^2}+\lambda_2)<r^{1/n}_{-}(\sqrt{1+\lambda_1^2}+\lambda_1)$ \ie when branes intersect the refined \Renyi entropy is negative.

\section{Discussion}
In this paper, we revisit the calculation of \Renyi entropy in AdS$_3$/(B)CFT$_2$ and we find that there is a connection between the negativity of \Renyi entropy and brane intersection. In some cases \cite{Biswas:2022xfw,Miyaji:2022cma}, gravity solutions with brane intersection seem to be physical because they give the correct spectrum of boundary-condition-changing operators in BCFT. In some cases \cite{Cooper:2018cmb,Geng:2021iyq,Kusuki:2022ozk}, the brane (self-)intersection is not physical because the energy of the corresponding state is beyond the black hole threshold. Our results also suggest that the brane intersection should be regularized or the semi-classical gravity theory is not the full theory. Of course, our result is far away from a proof. Following \cite{Kusuki:2022ozk},  more rigorous results may be obtained from a bootstrap perspective. 

We also explain the origin of the cut-off proposal \cite{Kusuki:2022ozk} given by the EOW branes. We check it in both the BTZ geometry and \Banados geometry. Applying Dong's formula we also obtain the finite part of the \Renyi entropy, the boundary entropy. When the brane intersection happens, Dong's formula also gives a negative result since the orientation of the spacetime is changed. We believe that similar results can be found in the higher dimensional cases. Another direction of generalization is to consider multiple intervals. 

We show that the replica partition function can be obtained by considering a \Banados geometry with EOW branes. The crucial point is that in the \Banados geometry, the EOW brane profile should be modified accordingly. In \cite{Caputa:2022zsr}, a similar geometry is proposed to dual the Virasoro coherent state. So our results may help to study such states.

\begin{figure}[h!]
\centering
        \includegraphics[scale=0.3]{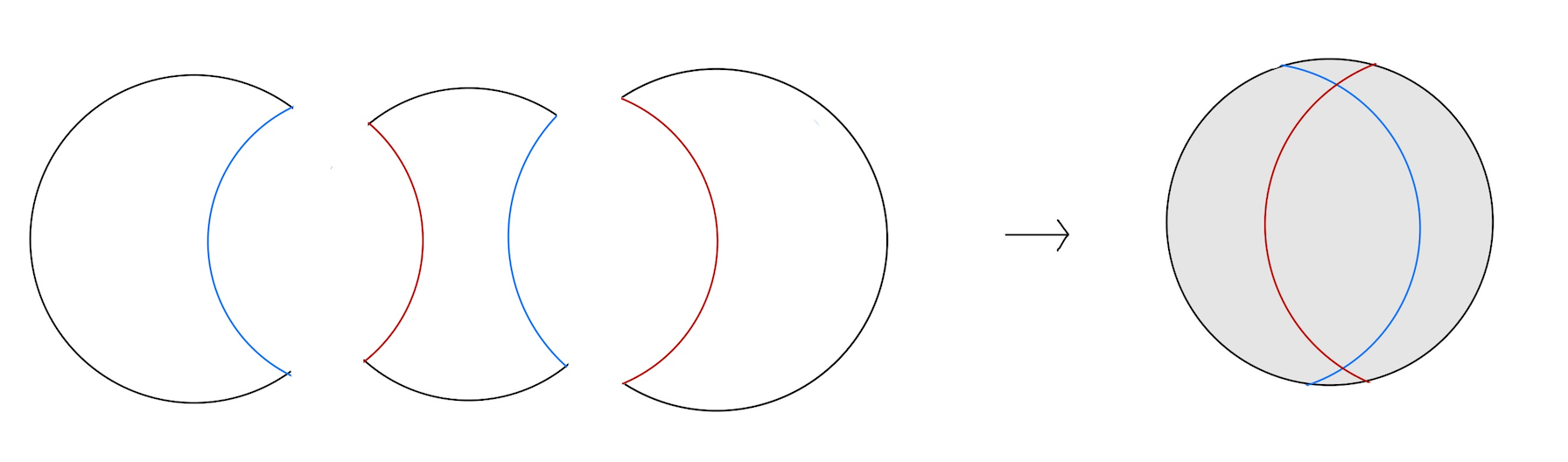}
  \caption{The figure is essentially the same one as figure 5 in \cite{Bah:2022uyz}. Here the blue and red boundaries are EOW branes and the EOW branes with the same color are glued together.}\label{cutglue}
\end{figure}

The geometries with brane intersection can also be constructed from a cut-and-glue procedure \cite{Bah:2022uyz} (see also \cite{Chandra:2022fwi}) as shown in Fig.\ref{cutglue}. The orientation of the central region gets inverted. However, it is argued in \cite{Bah:2022uyz} that this ill-defined geometry can be cured by including a suitable period of Lorentzian evolution. 
It is interesting to explore whether the geometries which are studied in this paper can be regularized in a similar way.

In a recent work \cite{Kanda:2023zse}, AdS/BCFT is generalized to include a brane-localized scalar that can describe non-conformal boundary conditions. Interestingly by adding the scalar degree of freedoms the EOW brane can have a large number of new solutions such as the connected EOW solution in the \Poincare metric which is not possible without the brane-localized scalar field. One important application of this setup is to study the boundary RG flow. Similar to the field theory arguments \cite{Friedan:2005dj,Friedan:2012jk} it is possible to use this setup to show that boundary entropy has a lower bound from a holographic point of view such that the brane intersection never happens.

\section*{Acknowledgments}
%	\emph{Acknowledgments:} 
We thank Cheng Peng for the valuable discussion and comments on an early version of the draft.  We thank many of the members of KITS for interesting related discussions. 
JT is supported by  the National Youth Fund No.12105289 and funds from the UCAS program of special research associate. XX is supported by NSFC NO. 12175237 and the Fundamental Research Funds for the Central Universities.


\begin{thebibliography}{50}
\bibitem{renyi}
A.~Rényi, \emph{On measures of entropy and information},  in \emph{Proceedings
  of the Fourth Berkeley Symposium on Mathematical Statistics and Probability,
  Volume 1: Contributions to the Theory of Statistics}, pp.~547--561,
  University of California Press, 1961,
  \href{http://projecteuclid.org/euclid.bsmsp/1200512181}{http://projecteuclid%
.org/euclid.bsmsp/1200512181}.
%%%%Renyi entropy
%\cite{Franchini:2007eu}
\bibitem{Franchini:2007eu}
F.~Franchini, A.~R.~Its and V.~E.~Korepin,
``Renyi Entropy of the XY Spin Chain,''
J. Phys. A \textbf{41}, 025302 (2008)
doi:10.1088/1751-8113/41/2/025302
[arXiv:0707.2534 [quant-ph]].
%52 citations counted in INSPIRE as of 20 Feb 2023
%\cite{Hayden:2016cfa}
\bibitem{Hayden:2016cfa}
P.~Hayden, S.~Nezami, X.~L.~Qi, N.~Thomas, M.~Walter and Z.~Yang,
``Holographic duality from random tensor networks,''
JHEP \textbf{11}, 009 (2016)
doi:10.1007/JHEP11(2016)009
[arXiv:1601.01694 [hep-th]].
%\cite{Klebanov:2011uf}
\bibitem{Klebanov:2011uf}
I.~R.~Klebanov, S.~S.~Pufu, S.~Sachdev and B.~R.~Safdi,
``Renyi Entropies for Free Field Theories,''
JHEP \textbf{04}, 074 (2012)
doi:10.1007/JHEP04(2012)074
[arXiv:1111.6290 [hep-th]].
%115 citations counted in INSPIRE as of 20 Feb 2023
%\cite{Holzhey:1994we}
\bibitem{Holzhey:1994we}
C.~Holzhey, F.~Larsen and F.~Wilczek,
``Geometric and renormalized entropy in conformal field theory,''
Nucl. Phys. B \textbf{424}, 443-467 (1994)
doi:10.1016/0550-3213(94)90402-2
[arXiv:hep-th/9403108 [hep-th]].
%1183 citations counted in INSPIRE as of 20 Feb 2023
%\cite{Calabrese:2009ez}
\bibitem{Calabrese:2009ez}
P.~Calabrese, J.~Cardy and E.~Tonni,
``Entanglement entropy of two disjoint intervals in conformal field theory,''
J. Stat. Mech. \textbf{0911}, P11001 (2009)
doi:10.1088/1742-5468/2009/11/P11001
[arXiv:0905.2069 [hep-th]].
%268 citations counted in INSPIRE as of 20 Feb 2023
%\cite{Hartman:2013mia}
\bibitem{Hartman:2013mia}
T.~Hartman,
``Entanglement Entropy at Large Central Charge,''
[arXiv:1303.6955 [hep-th]].
%375 citations counted in INSPIRE as of 20 Feb 2023
%\cite{Chen:2013kpa}
\bibitem{Chen:2013kpa}
B.~Chen and J.~J.~Zhang,
``On short interval expansion of R\'enyi entropy,''
JHEP \textbf{11}, 164 (2013)
doi:10.1007/JHEP11(2013)164
[arXiv:1309.5453 [hep-th]].
%78 citations counted in INSPIRE as of 20 Feb 2023
%\cite{Datta:2013hba}
\bibitem{Datta:2013hba}
S.~Datta and J.~R.~David,
``R\'enyi entropies of free bosons on the torus and holography,''
JHEP \textbf{04}, 081 (2014)
doi:10.1007/JHEP04(2014)081
[arXiv:1311.1218 [hep-th]].
%46 citations counted in INSPIRE as of 20 Feb 2023
%\cite{Perlmutter:2013paa}
\bibitem{Perlmutter:2013paa}
E.~Perlmutter,
``Comments on Renyi entropy in AdS$_3$/CFT$_2$,''
JHEP \textbf{05}, 052 (2014)
doi:10.1007/JHEP05(2014)052
[arXiv:1312.5740 [hep-th]].
%55 citations counted in INSPIRE as of 20 Feb 2023
%\cite{Perlmutter:2015iya}
\bibitem{Perlmutter:2015iya}
E.~Perlmutter,
``Virasoro conformal blocks in closed form,''
JHEP \textbf{08}, 088 (2015)
doi:10.1007/JHEP08(2015)088
[arXiv:1502.07742 [hep-th]].
%93 citations counted in INSPIRE as of 20 Feb 2023
%\cite{Headrick:2015gba}
\bibitem{Headrick:2015gba}
M.~Headrick, A.~Maloney, E.~Perlmutter and I.~G.~Zadeh,
``R\'enyi entropies, the analytic bootstrap, and 3D quantum gravity at higher genus,''
JHEP \textbf{07}, 059 (2015)
doi:10.1007/JHEP07(2015)059
[arXiv:1503.07111 [hep-th]].
%51 citations counted in INSPIRE as of 20 Feb 2023
%\cite{Perlmutter:2013gua}
\bibitem{Perlmutter:2013gua}
E.~Perlmutter,
``A universal feature of CFT R\'enyi entropy,''
JHEP \textbf{03}, 117 (2014)
doi:10.1007/JHEP03(2014)117
[arXiv:1308.1083 [hep-th]].
%83 citations counted in INSPIRE as of 20 Feb 2023
%\cite{Lee:2014xwa}
\bibitem{Lee:2014xwa}
J.~Lee, L.~McGough and B.~R.~Safdi,
``R\'enyi entropy and geometry,''
Phys. Rev. D \textbf{89}, no.12, 125016 (2014)
doi:10.1103/PhysRevD.89.125016
[arXiv:1403.1580 [hep-th]].
%32 citations counted in INSPIRE as of 20 Feb 2023
%\cite{Hung:2014npa}
\bibitem{Hung:2014npa}
L.~Y.~Hung, R.~C.~Myers and M.~Smolkin,
``Twist operators in higher dimensions,''
JHEP \textbf{10}, 178 (2014)
doi:10.1007/JHEP10(2014)178
[arXiv:1407.6429 [hep-th]].
%107 citations counted in INSPIRE as of 20 Feb 2023
%\cite{Allais:2014ata}
\bibitem{Allais:2014ata}
A.~Allais and M.~Mezei,
``Some results on the shape dependence of entanglement and R\'enyi entropies,''
Phys. Rev. D \textbf{91}, no.4, 046002 (2015)
doi:10.1103/PhysRevD.91.046002
[arXiv:1407.7249 [hep-th]].
%73 citations counted in INSPIRE as of 20 Feb 2023
%\cite{Lee:2014zaa}
\bibitem{Lee:2014zaa}
J.~Lee, A.~Lewkowycz, E.~Perlmutter and B.~R.~Safdi,
``R\'enyi entropy, stationarity, and entanglement of the conformal scalar,''
JHEP \textbf{03}, 075 (2015)
doi:10.1007/JHEP03(2015)075
[arXiv:1407.7816 [hep-th]].
%49 citations counted in INSPIRE as of 20 Feb 2023
%\cite{Lewkowycz:2014jia}
\bibitem{Lewkowycz:2014jia}
A.~Lewkowycz and E.~Perlmutter,
``Universality in the geometric dependence of Renyi entropy,''
JHEP \textbf{01}, 080 (2015)
doi:10.1007/JHEP01(2015)080
[arXiv:1407.8171 [hep-th]].
%69 citations counted in INSPIRE as of 20 Feb 2023
%\cite{Bueno:2015lza}
\bibitem{Bueno:2015lza}
P.~Bueno and R.~C.~Myers,
``Universal entanglement for higher dimensional cones,''
JHEP \textbf{12}, 168 (2015)
doi:10.1007/JHEP12(2015)168
[arXiv:1508.00587 [hep-th]].
%50 citations counted in INSPIRE as of 20 Feb 2023
%\cite{Bianchi:2015liz}
\bibitem{Bianchi:2015liz}
L.~Bianchi, M.~Meineri, R.~C.~Myers and M.~Smolkin,
``R\'enyi entropy and conformal defects,''
JHEP \textbf{07}, 076 (2016)
doi:10.1007/JHEP07(2016)076
[arXiv:1511.06713 [hep-th]].
%91 citations counted in INSPIRE as of 20 Feb 2023
%\cite{Dong:2016wcf}
\bibitem{Dong:2016wcf}
X.~Dong,
``Shape Dependence of Holographic R\'enyi Entropy in Conformal Field Theories,''
Phys. Rev. Lett. \textbf{116}, no.25, 251602 (2016)
doi:10.1103/PhysRevLett.116.251602
[arXiv:1602.08493 [hep-th]].
%47 citations counted in INSPIRE as of 20 Feb 2023
%\cite{Headrick:2010zt}
\bibitem{Headrick:2010zt}
M.~Headrick,
``Entanglement Renyi entropies in holographic theories,''
Phys. Rev. D \textbf{82}, 126010 (2010)
doi:10.1103/PhysRevD.82.126010
[arXiv:1006.0047 [hep-th]].
%519 citations counted in INSPIRE as of 20 Feb 2023
%\cite{Hung:2011nu}
\bibitem{Hung:2011nu}
L.~Y.~Hung, R.~C.~Myers, M.~Smolkin and A.~Yale,
``Holographic Calculations of Renyi Entropy,''
JHEP \textbf{12}, 047 (2011)
doi:10.1007/JHEP12(2011)047
[arXiv:1110.1084 [hep-th]].
%205 citations counted in INSPIRE as of 06 Feb 2023
%\cite{Fursaev:2012mp}
\bibitem{Fursaev:2012mp}
D.~V.~Fursaev,
``Entanglement Renyi Entropies in Conformal Field Theories and Holography,''
JHEP \textbf{05}, 080 (2012)
doi:10.1007/JHEP05(2012)080
[arXiv:1201.1702 [hep-th]].
%62 citations counted in INSPIRE as of 20 Feb 2023
%\cite{Faulkner:2013yia}
\bibitem{Faulkner:2013yia}
T.~Faulkner,
``The Entanglement Renyi Entropies of Disjoint Intervals in AdS/CFT,''
[arXiv:1303.7221 [hep-th]].
%256 citations counted in INSPIRE as of 20 Feb 2023
%\cite{Galante:2013wta}
\bibitem{Galante:2013wta}
D.~A.~Galante and R.~C.~Myers,
``Holographic Renyi entropies at finite coupling,''
JHEP \textbf{08}, 063 (2013)
doi:10.1007/JHEP08(2013)063
[arXiv:1305.7191 [hep-th]].
%31 citations counted in INSPIRE as of 20 Feb 2023
%\cite{Belin:2013dva}
\bibitem{Belin:2013dva}
A.~Belin, A.~Maloney and S.~Matsuura,
``Holographic Phases of Renyi Entropies,''
JHEP \textbf{12}, 050 (2013)
doi:10.1007/JHEP12(2013)050
[arXiv:1306.2640 [hep-th]].
%63 citations counted in INSPIRE as of 20 Feb 2023
%\cite{Barrella:2013wja}
\bibitem{Barrella:2013wja}
T.~Barrella, X.~Dong, S.~A.~Hartnoll and V.~L.~Martin,
``Holographic entanglement beyond classical gravity,''
JHEP \textbf{09}, 109 (2013)
doi:10.1007/JHEP09(2013)109
[arXiv:1306.4682 [hep-th]].
%181 citations counted in INSPIRE as of 20 Feb 2023
%\cite{Chen:2013dxa}
\bibitem{Chen:2013dxa}
B.~Chen, J.~Long and J.~j.~Zhang,
``Holographic R\'enyi entropy for CFT with W symmetry,''
JHEP \textbf{04}, 041 (2014)
doi:10.1007/JHEP04(2014)041
[arXiv:1312.5510 [hep-th]].
%67 citations counted in INSPIRE as of 20 Feb 2023
%\cite{Belin:2013uta}
\bibitem{Belin:2013uta}
A.~Belin, L.~Y.~Hung, A.~Maloney, S.~Matsuura, R.~C.~Myers and T.~Sierens,
``Holographic Charged Renyi Entropies,''
JHEP \textbf{12}, 059 (2013)
doi:10.1007/JHEP12(2013)059
[arXiv:1310.4180 [hep-th]].
%124 citations counted in INSPIRE as of 20 Feb 2023
%\cite{Nishioka:2013haa}
\bibitem{Nishioka:2013haa}
T.~Nishioka and I.~Yaakov,
``Supersymmetric Renyi Entropy,''
JHEP \textbf{10}, 155 (2013)
doi:10.1007/JHEP10(2013)155
[arXiv:1306.2958 [hep-th]].
%86 citations counted in INSPIRE as of 20 Feb 2023
%\cite{Alday:2014fsa}
\bibitem{Alday:2014fsa}
L.~F.~Alday, P.~Richmond and J.~Sparks,
``The holographic supersymmetric Renyi entropy in five dimensions,''
JHEP \textbf{02}, 102 (2015)
doi:10.1007/JHEP02(2015)102
[arXiv:1410.0899 [hep-th]].
%34 citations counted in INSPIRE as of 20 Feb 2023
%\cite{Giveon:2015cgs}
\bibitem{Giveon:2015cgs}
A.~Giveon and D.~Kutasov,
``Supersymmetric Renyi entropy in CFT$_{2}$ and AdS$_{3}$,''
JHEP \textbf{01}, 042 (2016)
doi:10.1007/JHEP01(2016)042
[arXiv:1510.08872 [hep-th]].
%23 citations counted in INSPIRE as of 20 Feb 2023
%\cite{Chu:2016tps}
\bibitem{Chu:2016tps}
C.~S.~Chu and R.~X.~Miao,
``Universality in the shape dependence of holographic R\'enyi entropy for general higher derivative gravity,''
JHEP \textbf{12}, 036 (2016)
doi:10.1007/JHEP12(2016)036
[arXiv:1608.00328 [hep-th]].
%29 citations counted in INSPIRE as of 20 Feb 2023
%\cite{Dey:2016pei}
\bibitem{Dey:2016pei}
A.~Dey, P.~Roy and T.~Sarkar,
``On holographic R\'enyi entropy in some modified theories of gravity,''
JHEP \textbf{04}, 098 (2018)
doi:10.1007/JHEP04(2018)098
[arXiv:1609.02290 [hep-th]].
%24 citations counted in INSPIRE as of 20 Feb 2023
%\cite{Belin:2017nze}
\bibitem{Belin:2017nze}
A.~Belin, C.~A.~Keller and I.~G.~Zadeh,
``Genus two partition functions and R\'enyi entropies of large c conformal field theories,''
J. Phys. A \textbf{50}, no.43, 435401 (2017)
doi:10.1088/1751-8121/aa8a11
[arXiv:1704.08250 [hep-th]].
%45 citations counted in INSPIRE as of 20 Feb 2023
%\cite{Jiang:2017ecm}
\bibitem{Jiang:2017ecm}
H.~Jiang, W.~Song and Q.~Wen,
``Entanglement Entropy in Flat Holography,''
JHEP \textbf{07}, 142 (2017)
doi:10.1007/JHEP07(2017)142
[arXiv:1706.07552 [hep-th]].
%73 citations counted in INSPIRE as of 20 Feb 2023
%\cite{Dong:2018cuv}
\bibitem{Dong:2018cuv}
X.~Dong, E.~Silverstein and G.~Torroba,
``De Sitter Holography and Entanglement Entropy,''
JHEP \textbf{07}, 050 (2018)
doi:10.1007/JHEP07(2018)050
[arXiv:1804.08623 [hep-th]].
%80 citations counted in INSPIRE as of 20 Feb 2023
%\cite{Donnelly:2018bef}
\bibitem{Donnelly:2018bef}
W.~Donnelly and V.~Shyam,
``Entanglement entropy and $T \overline{T}$ deformation,''
Phys. Rev. Lett. \textbf{121}, no.13, 131602 (2018)
doi:10.1103/PhysRevLett.121.131602
[arXiv:1806.07444 [hep-th]].
%92 citations counted in INSPIRE as of 20 Feb 2023
%\cite{Dong:2018lsk}
\bibitem{Dong:2018lsk}
X.~Dong,
``Holographic R\'enyi Entropy at High Energy Density,''
Phys. Rev. Lett. \textbf{122}, no.4, 041602 (2019)
doi:10.1103/PhysRevLett.122.041602
[arXiv:1811.04081 [hep-th]].
%14 citations counted in INSPIRE as of 20 Feb 2023
%\cite{Akers:2018fow}
\bibitem{Akers:2018fow}
C.~Akers and P.~Rath,
``Holographic Renyi Entropy from Quantum Error Correction,''
JHEP \textbf{05}, 052 (2019)
doi:10.1007/JHEP05(2019)052
[arXiv:1811.05171 [hep-th]].
%69 citations counted in INSPIRE as of 20 Feb 2023
%\cite{Rabenstein:2018bri}
\bibitem{Rabenstein:2018bri}
A.~Rabenstein, N.~Bodendorfer, P.~Buividovich and A.~Sch\"afer,
``Lattice study of R\'enyi entanglement entropy in $SU(N_c)$ lattice Yang-Mills theory with $N_c = 2, 3, 4$,''
Phys. Rev. D \textbf{100}, no.3, 034504 (2019)
doi:10.1103/PhysRevD.100.034504
[arXiv:1812.04279 [hep-lat]].
%18 citations counted in INSPIRE as of 20 Feb 2023
%\cite{Jeong:2019ylz}
\bibitem{Jeong:2019ylz}
H.~S.~Jeong, K.~Y.~Kim and M.~Nishida,
``Entanglement and R\'enyi entropy of multiple intervals in $T\overline{T}$-deformed CFT and holography,''
Phys. Rev. D \textbf{100}, no.10, 106015 (2019)
doi:10.1103/PhysRevD.100.106015
[arXiv:1906.03894 [hep-th]].
%30 citations counted in INSPIRE as of 20 Feb 2023
%\cite{Botta-Cantcheff:2020ywu}
\bibitem{Botta-Cantcheff:2020ywu}
M.~Botta-Cantcheff, P.~J.~Martinez and J.~F.~Zarate,
``R\'enyi entropies and area operator from gravity with Hayward term,''
JHEP \textbf{07}, no.07, 227 (2020)
doi:10.1007/JHEP07(2020)227
[arXiv:2005.11338 [hep-th]].
%5 citations counted in INSPIRE as of 20 Feb 2023
%\cite{Dong:2020iod}
\bibitem{Dong:2020iod}
X.~Dong and H.~Wang,
``Enhanced corrections near holographic entanglement transitions: a chaotic case study,''
JHEP \textbf{11}, 007 (2020)
doi:10.1007/JHEP11(2020)007
[arXiv:2006.10051 [hep-th]].
%33 citations counted in INSPIRE as of 20 Feb 2023
%\cite{Dong:2020uxp}
\bibitem{Dong:2020uxp}
X.~Dong, X.~L.~Qi, Z.~Shangnan and Z.~Yang,
``Effective entropy of quantum fields coupled with gravity,''
JHEP \textbf{10}, 052 (2020)
doi:10.1007/JHEP10(2020)052
[arXiv:2007.02987 [hep-th]].
%58 citations counted in INSPIRE as of 20 Feb 2023
%%%%%Renyi entropy
%383 citations counted in INSPIRE as of 20 Feb 2023
%\cite{Bai:2022obp}
\bibitem{Bai:2022obp}
X.~Bai and J.~Ren,
``Holographic R\'enyi entropies from hyperbolic black holes with scalar hair,''
JHEP \textbf{12}, 038 (2022)
doi:10.1007/JHEP12(2022)038
[arXiv:2210.03732 [hep-th]].
%2 citations counted in INSPIRE as of 20 Feb 2023
%\cite{Affleck:1991tk}
\bibitem{Affleck:1991tk}
I.~Affleck and A.~W.~W.~Ludwig,
``Universal noninteger 'ground state degeneracy' in critical quantum systems,''
Phys. Rev. Lett. \textbf{67}, 161-164 (1991)
doi:10.1103/PhysRevLett.67.161
%616 citations counted in INSPIRE as of 10 Feb 2023
%\cite{Azeyanagi:2007qj}
\bibitem{Azeyanagi:2007qj}
T.~Azeyanagi, A.~Karch, T.~Takayanagi and E.~G.~Thompson,
``Holographic calculation of boundary entropy,''
JHEP \textbf{03}, 054 (2008)
doi:10.1088/1126-6708/2008/03/054
[arXiv:0712.1850 [hep-th]].
%98 citations counted in INSPIRE as of 20 Feb 2023
%\cite{Cardy:2016fqc}
\bibitem{Cardy:2016fqc}
J.~Cardy and E.~Tonni,
``Entanglement hamiltonians in two-dimensional conformal field theory,''
J. Stat. Mech. \textbf{1612}, no.12, 123103 (2016)
doi:10.1088/1742-5468/2016/12/123103
[arXiv:1608.01283 [cond-mat.stat-mech]].
%181 citations counted in INSPIRE as of 03 Jan 2023
%\cite{DiFrancesco:1997nk}
\bibitem{DiFrancesco:1997nk}
P.~Di Francesco, P.~Mathieu and D.~Senechal,
``Conformal Field Theory,''
Springer-Verlag, 1997,
ISBN 978-0-387-94785-3, 978-1-4612-7475-9
doi:10.1007/978-1-4612-2256-9
%487 citations counted in INSPIRE as of 20 Feb 2023
%\cite{Karch:2000gx}
%\cite{Friedan:2005dj}
\bibitem{Friedan:2005dj}
D.~Friedan and A.~Konechny,
``Infrared properties of boundaries in 1-d quantum systems,''
J. Stat. Mech. \textbf{0603}, P03014 (2006)
doi:10.1088/1742-5468/2006/03/P03014
[arXiv:hep-th/0512023 [hep-th]].
%7 citations counted in INSPIRE as of 20 Feb 2023
%\cite{Friedan:2012jk}
\bibitem{Friedan:2012jk}
D.~Friedan, A.~Konechny and C.~Schmidt-Colinet,
``Lower bound on the entropy of boundaries and junctions in 1+1d quantum critical systems,''
Phys. Rev. Lett. \textbf{109}, 140401 (2012)
doi:10.1103/PhysRevLett.109.140401
[arXiv:1206.5395 [hep-th]].
%17 citations counted in INSPIRE as of 20 Feb 2023
\bibitem{Karch:2000gx}
A.~Karch and L.~Randall,
``Open and closed string interpretation of SUSY CFT's on branes with boundaries,''
JHEP \textbf{06}, 063 (2001)
doi:10.1088/1126-6708/2001/06/063
[arXiv:hep-th/0105132 [hep-th]].
%480 citations counted in INSPIRE as of 10 Feb 2023
%\cite{Takayanagi:2011zk}
\bibitem{Takayanagi:2011zk}
T.~Takayanagi,
``Holographic Dual of BCFT,''
Phys. Rev. Lett. \textbf{107}, 101602 (2011)
doi:10.1103/PhysRevLett.107.101602
[arXiv:1105.5165 [hep-th]].
%354 citations counted in INSPIRE as of 10 Feb 2023
\bibitem{Kusuki:2022ozk}
Y.~Kusuki and Z.~Wei,
``AdS/BCFT from Conformal Bootstrap: Construction of Gravity with Branes and Particles,''
[arXiv:2210.03107 [hep-th]].
%2 citations counted in INSPIRE as of 03 Jan 2023
\bibitem{Miyaji:2022cma}
M.~Miyaji and C.~Murdia,
``Holographic BCFT with a Defect on the End-of-the-World Brane,''
[arXiv:2208.13783 [hep-th]].
%\cite{Biswas:2022xfw}
\bibitem{Biswas:2022xfw}
S.~Biswas, J.~Kastikainen, S.~Shashi and J.~Sully,
``Holographic BCFT spectra from brane mergers,''
JHEP \textbf{11}, 158 (2022)
doi:10.1007/JHEP11(2022)158
[arXiv:2209.11227 [hep-th]].
%3 citations counted in INSPIRE as of 03 Jan 2023
%\cite{Kontsevich:2021dmb}
\bibitem{Kontsevich:2021dmb}
M.~Kontsevich and G.~Segal,
%``Wick Rotation and the Positivity of Energy in Quantum Field Theory,''
Quart. J. Math. Oxford Ser. \textbf{72}, no.1-2, 673-699 (2021)
doi:10.1093/qmath/haab027
[arXiv:2105.10161 [hep-th]].
%50 citations counted in INSPIRE as of 20 Feb 2023
%\cite{Witten:2021nzp}
\bibitem{Witten:2021nzp}
E.~Witten,
``A Note On Complex Spacetime Metrics,''
[arXiv:2111.06514 [hep-th]].
%64 citations counted in INSPIRE as of 20 Feb 2023
%\cite{Bah:2022uyz}
\bibitem{Bah:2022uyz}
I.~Bah, Y.~Chen and J.~Maldacena,
``Estimating global charge violating amplitudes from wormholes,''
[arXiv:2212.08668 [hep-th]].
%2 citations counted in INSPIRE as of 20 Feb 2023

%\cite{Dong:2016fnf}
\bibitem{Dong:2016fnf}
X.~Dong,
``The Gravity Dual of Renyi Entropy,''
Nature Commun. \textbf{7}, 12472 (2016)
doi:10.1038/ncomms12472
[arXiv:1601.06788 [hep-th]].
%224 citations counted in INSPIRE as of 03 Jan 2023


\bibitem{Calabrese:2004eu}
P.~Calabrese and J.~L.~Cardy,
``Entanglement entropy and quantum field theory,''
J. Stat. Mech. \textbf{0406}, P06002 (2004)
doi:10.1088/1742-5468/2004/06/P06002
[arXiv:hep-th/0405152 [hep-th]].
%1755 citations counted in INSPIRE as of 05 Feb 2023
%\cite{Lunin:2000yv}
\bibitem{Lunin:2000yv}
O.~Lunin and S.~D.~Mathur,
``Correlation functions for M**N / S(N) orbifolds,''
Commun. Math. Phys. \textbf{219}, 399-442 (2001)
doi:10.1007/s002200100431
[arXiv:hep-th/0006196 [hep-th]].

\bibitem{Herzog:2015ioa}
C.~P.~Herzog, K.~W.~Huang and K.~Jensen,
``Universal Entanglement and Boundary Geometry in Conformal Field Theory,''
JHEP \textbf{01}, 162 (2016)
doi:10.1007/JHEP01(2016)162
[arXiv:1510.00021 [hep-th]].
%71 citations counted in INSPIRE as of 03 Jan 2023
\bibitem{Liouville}
D. Friedan, ``Introduction To Polyakov's String Theory," in \textit{Recent Advances in Field Theory and Statistical Mechanics}, ed. by J. B. Zuber and R. Stora. North-Holland, 1984.
\bibitem{Hayward:1993my}
G.~Hayward,
``Gravitational action for space-times with nonsmooth boundaries,''
Phys. Rev. D \textbf{47}, 3275-3280 (1993)
doi:10.1103/PhysRevD.47.3275

\bibitem{Fujita:2011fp}
M.~Fujita, T.~Takayanagi and E.~Tonni,
``Aspects of AdS/BCFT,''
JHEP \textbf{11}, 043 (2011)
doi:10.1007/JHEP11(2011)043
[arXiv:1108.5152 [hep-th]].
%\cite{Wald:1993nt}
\bibitem{Wald:1993nt}
R.~M.~Wald,
``Black hole entropy is the Noether charge,''
Phys. Rev. D \textbf{48}, no.8, R3427-R3431 (1993)
doi:10.1103/PhysRevD.48.R3427
[arXiv:gr-qc/9307038 [gr-qc]].
%2006 citations counted in INSPIRE as of 10 Feb 2023
%\cite{Roberts:2012aq}
%\cite{Cooper:2018cmb}
\bibitem{Cooper:2018cmb}
S.~Cooper, M.~Rozali, B.~Swingle, M.~Van Raamsdonk, C.~Waddell and D.~Wakeham,
``Black hole microstate cosmology,''
JHEP \textbf{07}, 065 (2019)
doi:10.1007/JHEP07(2019)065
[arXiv:1810.10601 [hep-th]].
%93 citations counted in INSPIRE as of 10 Feb 2023
%\cite{Geng:2021iyq}
\bibitem{Geng:2021iyq}
H.~Geng, S.~L\"ust, R.~K.~Mishra and D.~Wakeham,
``Holographic BCFTs and Communicating Black Holes,''
jhep \textbf{08}, 003 (2021)
doi:10.1007/JHEP08(2021)003
[arXiv:2104.07039 [hep-th]].
%65 citations counted in INSPIRE as of 10 Feb 2023
%\cite{Bianchi:2022ulu}
\bibitem{Bianchi:2022ulu}
L.~Bianchi, S.~De Angelis and M.~Meineri,
``Radiation, entanglement and islands from a boundary local quench,''
[arXiv:2203.10103 [hep-th]].
%11 citations counted in INSPIRE as of 10 Feb 2023
%\cite{Kawamoto:2022etl}
\bibitem{Kawamoto:2022etl}
T.~Kawamoto, T.~Mori, Y.~k.~Suzuki, T.~Takayanagi and T.~Ugajin,
``Holographic local operator quenches in BCFTs,''
JHEP \textbf{05}, 060 (2022)
doi:10.1007/JHEP05(2022)060
[arXiv:2203.03851 [hep-th]].
%11 citations counted in INSPIRE as of 10 Feb 2023
%\cite{Banados:1998gg}
\bibitem{Banados:1998gg}
M.~Banados,
``Three-dimensional quantum geometry and black holes,''
AIP Conf. Proc. \textbf{484}, no.1, 147-169 (1999)
doi:10.1063/1.59661
[arXiv:hep-th/9901148 [hep-th]].
%338 citations counted in INSPIRE as of 11 Feb 2023
\bibitem{Roberts:2012aq}
M.~M.~Roberts,
``Time evolution of entanglement entropy from a pulse,''
JHEP \textbf{12}, 027 (2012)
doi:10.1007/JHEP12(2012)027
[arXiv:1204.1982 [hep-th]].
%79 citations counted in INSPIRE as of 03 Dec 2022
%\cite{Skenderis:1999nb}
\bibitem{Skenderis:1999nb}
K.~Skenderis and S.~N.~Solodukhin,
``Quantum effective action from the AdS / CFT correspondence,''
Phys. Lett. B \textbf{472}, 316-322 (2000)
doi:10.1016/S0370-2693(99)01467-7
[arXiv:hep-th/9910023 [hep-th]].
%200 citations counted in INSPIRE as of 06 Feb 2023
%\cite{Sully:2020pza}
\bibitem{Sully:2020pza}
J.~Sully, M.~Van Raamsdonk and D.~Wakeham,
``BCFT entanglement entropy at large central charge and the black hole interior,''
JHEP \textbf{03}, 167 (2021)
doi:10.1007/JHEP03(2021)167
[arXiv:2004.13088 [hep-th]].
%80 citations counted in INSPIRE as of 12 Feb 2023
%\cite{Caputa:2022zsr}
\bibitem{Caputa:2022zsr}
P.~Caputa and D.~Ge,
``Entanglement and geometry from subalgebras of the Virasoro,''
[arXiv:2211.03630 [hep-th]].
%5 citations counted in INSPIRE as of 12 Feb 2023
%\cite{Chandra:2022fwi}
\bibitem{Chandra:2022fwi}
J.~Chandra and T.~Hartman,
``Coarse graining pure states in AdS/CFT,''
[arXiv:2206.03414 [hep-th]].
%12 citations counted in INSPIRE as of 20 Feb 2023
%\cite{Kanda:2023zse}
\bibitem{Kanda:2023zse}
H.~Kanda, M.~Sato, Y.~k.~Suzuki, T.~Takayanagi and Z.~Wei,
``AdS/BCFT with Brane-Localized Scalar Field,''
[arXiv:2302.03895 [hep-th]].
%0 citations counted in INSPIRE as of 20 Feb 2023
\end{thebibliography}
\end{document}